\documentclass[prb,twocolumn,showpacs,superscriptaddress,preprintnumbers,floatfix,amsmath,amssymb]{revtex4-2}

\usepackage{amssymb}
\usepackage{graphicx}
\usepackage{dcolumn}
\usepackage{bm}
\usepackage{color}
\hfuzz=\maxdimen
\begin{document}

\title{Magnetic and superconducting phase diagram of Eu(Fe$_{1-x}$Ni$_{x}$)As$_{2}$}

\author{Ya-Bin Liu}
\affiliation{Department of Physics, Zhejiang University, Hangzhou 310027, China}

\author{Yi Liu}
\affiliation{Department of Physics, Zhejiang University, Hangzhou 310027, China}
\affiliation{Department of Applied Physics, Zhejiang University of Technology, Hangzhou 310023, China}

\author{Yan-Wei Cui}
\affiliation{Department of Physics, Zhejiang University, Hangzhou 310027, China}
\affiliation{School of Sciences, Westlake Institute for Advanced Study, Westlake University, Hangzhou 310064, China}

\author{Si-Qi Wu}
\affiliation{Department of Physics, Zhejiang University, Hangzhou 310027, China}
\author{Zhi Ren}
\affiliation{School of Sciences, Westlake Institute for Advanced Study, Westlake University, Hangzhou 310064, China}

\author{Guang-Han Cao} \email[]{ghcao@zju.edu.cn}
\affiliation{Department of Physics, Zhejiang University, Hangzhou 310027, China}
\affiliation{Zhejiang Province Key Laboratory of Quantum Technology and Devices, Interdisciplinary Center for Quantum Information, and State Key Lab of Silicon Materials, Zhejiang University, Hangzhou 310027, China}
\affiliation{Collaborative Innovation Centre of Advanced Microstructures, Nanjing University, Nanjing 210093, China}

\date{\today}

\begin{abstract}
We report the Ni-doping effect on magnetism and superconductivity (SC) in an Eu-containing 112-type system Eu(Fe$_{1-x}$Ni$_{x})$As$_{2}$ ($0\leq x\leq 0.15$) by the measurements of resistivity, magnetization, and specific heat. The undoped EuFeAs$_2$ undergoes a spin-density-wave (SDW) transition at $T_\mathrm{SDW}\sim$ 105 K in the Fe sublattice and a magnetic ordering at $T_\mathrm{m}\sim$ 40 K in the Eu sublattice. Complex Eu-spin magnetism is manifested by a spin-glass reentrance at $T_\mathrm{SG}\sim$ 15 K and an additional spin reorientation at $T_\mathrm{SR}\sim$ 7 K. With Ni doping, the SDW order is rapidly suppressed, and SC emerges in the Ni-doping range of 0.01 $\leq x\leq$ 0.1 where a maximum of the superconducting transition temperature $T_\mathrm{c}^{\mathrm{max}}=$ 17.6 K shows up at $x$ = 0.04. On the other hand, $T_\mathrm{m}$ decreases very slowly, yet $T_\mathrm{SG}$ and  $T_\mathrm{SR}$ hardly change with the Ni doping. The phase diagram has been established, which suggests a very weak coupling between SC and Eu spins. The complex Eu-spin magnetism is discussed in terms of the Ruderman-Kittel-Kasuya-Yosida interactions mediated by the conduction electrons from both layers of FeAs and As surrounding Eu$^{2+}$ ions.

\end{abstract}


\maketitle
\section{\label{sec:level1}Introduction}

Iron-based pnictide superconductors containing layers of Eu$^{2+}$ ions have received considerable attention primarily because of the coexistence of superconductivity (SC) and ferromagnetism (FM)~\cite{cao2012,dressel}. One typical example is the 122-type EuFe$_2$(As$_{1-x}$P$_{x}$)$_2$ which shows SC at $T_{\mathrm{c}}=$ 20-29 K and Eu$^{2+}$-spin FM at $T_{\mathrm{m}}\sim$ 18 K~\cite{ren2009,cao2011,Pogrebna2015} with spontaneous magnetization almost along the crystallographic $c$ axis~\cite{felner2011,nd2014,rxs2014}. A reentrant spin-glass transition was identified a few Kelvins below $T_{\mathrm{c}}$~\cite{zapf}. Another example is the 1144-type $A$EuFe$_4$As$_4$ ($A$ = Rb, Cs)~\cite{Eu1144,liuy2016-1,liuy2016-2}, where the Eu$^{2+}$ spins align within the $ab$ plane~\cite{Eu1144.Mossbauer,Eu1144.neutron}. Through Ni~\cite{ly.1144Ni} or Co~\cite{lyb.1144Co} doping, a rare superconducting ferromagnet with $T_{\mathrm{m}}>T_{\mathrm{c}}$ was realized. Interestingly, the interplay of between FM and SC generates new states of matter such as spontaneous vortex state~\cite{jiao2017,Eu1144.self-flux}, domain Meissner state, and vortex-antivortex state~\cite{Eu122P.DMS,Eu122P.SCFM}.

Recently, a new type of Eu-containing iron pnictide EuFeAs$_2$ was discovered~\cite{yu2017}. It is structurally analogous to the 112-type superconductors $A$FeAs$_2$ ($A=$ (Ca,La)~\cite{katayama2013} or (Ca,Pr)~\cite{yakita2014}) which can be stabilized only by the partial substitution with rare-earth elements~\cite{Sala2014}. Notably, the 112-type iron pnictides bear unique properties including unusual As$^-$ valence state~\cite{Ray2017,Nohara2017}, metallicity in the As plane~\cite{li2015.112,jiang2016.112}, and possible non-trivial topological state~\cite{wuxx.PRB2014,wuxx.prb2015,Dirac.APL2016,wuxx.PRX2020}, all of which are closely related to the zigzag As chains sandwiched by the $A$-site ions [see Fig.~\ref{xrd}(c)]. EuFeAs$_2$ is so far the unique 112 iron-pnictide parent compound that can be stabilized by itself. Measurements of electrical resistivity and magnetic susceptibility show two distinct anomalies at $T_{\mathrm{SDW}}\sim$ 110 K and $T_{\mathrm{m}}\sim$ 40 K, respectively~\cite{yu2017}. The M\"{o}ssbauer measurement demonstrated that the 110-K transition was due to a spin-density-wave (SDW) ordering in the Fe sublattice, and the 40-K transition was associated with a magnetic ordering in the Eu sublattice~\cite{albedah2020}. With the specific-heat and magnetization measurements, the present authors~\cite{liuyb2018} confirmed the dominant antiferromagnetic ordering of Eu spins at $T_{\mathrm{m}}$, and found a reentrant spin-glass transition at $T_\mathrm{SG}\sim$ 15 K and a spin reorientation at $T_\mathrm{SR}\sim$ 7 K. SC was realized with La doping~\cite{yu2017} or with Ni doping~\cite{liuyb2018}, and the $T_\mathrm{c}$ values achieve 11 K and 17.5 K in Eu$_{0.85}$La$_{0.15}$FeAs$_2$ and Eu(Fe$_{0.96}$Ni$_{0.04}$)As$_2$, respectively. In the case of Ni doping, a weak spontaneous magnetization coexisting with SC was observed at low temperatures.

To reveal the evolution of magnetism and SC with Ni doping, in this paper, we systematically study the physical properties of a series of 13 samples in Eu(Fe$_{1-x}$Ni$_{x}$)As$_{2}$ ($0\leq x\leq 0.15$). We find that the SDW order is quickly suppressed with the Ni doping, and it disappears at $x\geq$ 0.05. SC emerges in a broad Ni-doping range of 0.01 $\leq x\leq$ 0.1 with a dome-like $T_\mathrm{c}(x)$ dependence. On the other hand, the magnetic transitions associated with Eu$^{2+}$ spins changes slightly and subtly with the Ni doping. We propose that both the zigzag As chains and the FeAs layers mediate the Ruderman-Kittel-Kasuya-Yosida (RKKY) interactions, which jointly brings about the complex Eu-spin magnetism. The magnetic and superconducting phase diagram has been established, which suggests a very weak coupling between SC and Eu spins.

\section{\label{sec:level2}Experimental methods}

Polycrystalline samples of Eu(Fe$_{1-x}$Ni$_{x}$)As$_{2}$ ($x$ = 0, 0.005, 0.01, 0.02, 0.03, 0.04, 0.05, 0.06, 0.07, 0.08, 0.09, 0.1, and 0.15) were synthesized by high-temperature solid-state reactions in sealed silica ampoule using intermediate products of EuAs, FeAs, and NiAs. The intermediate products were prepared by reacting As pieces (99.999\%) with Eu pieces (99.9\%), Fe powders (99.998\%) and Ni powders (99.998\%), respectively, in sealed silica ampoules at 700-750$^{\circ}$C for 24 h. Mixtures of above binary arsenides in the stoichiometric composition of Eu(Fe$_{1-x}$Ni$_{x}$)As$_{2}$ were ground, pressed into pellets, and then sealed in evacuated silica ampoules. The sample-loaded ampoules were heated rapidly to 800-820$^{\circ}$C in a muffle furnace, holding for 30 h, followed by quenching. Additional sintering at the same temperature for one week was necessary for improving the quality of samples. Note that operations of weighing, grinding, pressing, and loading of samples were all carried out in a glove box filled with high-purity argon.

Powder x-ray diffraction (XRD) was carried out at room temperature using a PANalytical x-ray diffractometer (Model EMPYREAN) with monochromatic Cu-K$_{\alpha1}$ radiation. The lattice parameters were obtained by a least-squares fit of 15-25 reflections in the range of $5^{\circ}\leq 2\theta \leq 80^{\circ}$. The temperature dependence of electrical resistivity, ac susceptibility and heat capacity was measured on a Quantum Design Physical Property Measurement System (PPMS-9). In the resistance measurement, a four-electrode method was employed. The sample pellet was cut into a thin rectangular bar on which thin gold wires were attached with silver paint. For the ac susceptibility measurement, an oscillated field $H_\mathrm{ac}$ = 15 Oe was applied. The heat capacity was measured by a thermal relaxation method using sample plates with mass about 10 mg. The temperature dependence of dc magnetization under different magnetic fields was carried out on a Quantum Design Magnetic Property Measurement System (MPMS3). Measurement protocols with zero-field cooling (ZFC) and field cooling (FC) were employed at low magnetic fields.

\section{\label{sec:level3}Results and discussion}

\subsection{\label{subsec:level1}X-ray diffraction}

\begin{figure}
\includegraphics[width=8cm]{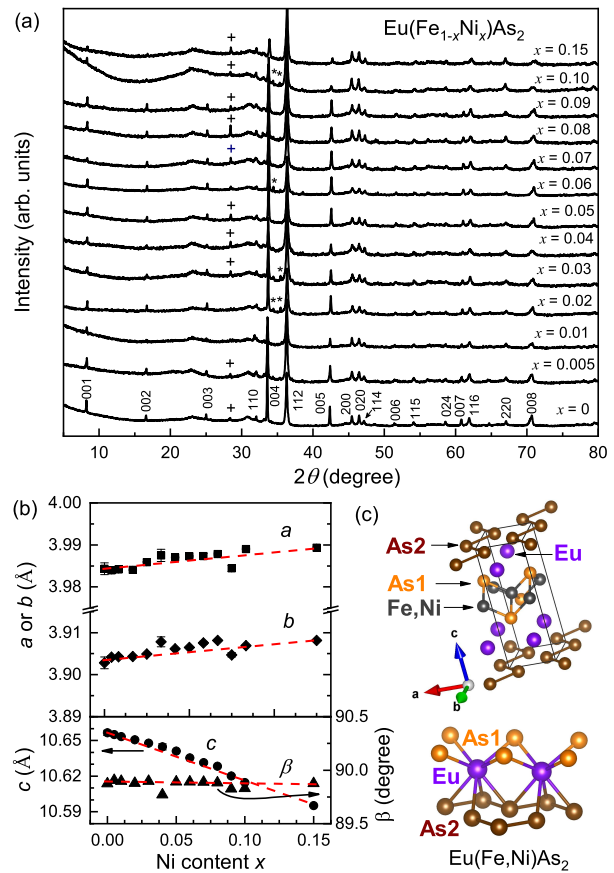}
\caption{(a) Powder x-ray diffraction patterns of Eu(Fe$_{1-x}$Ni$_{x})$As$_{2}$. Small unindexed peaks are identified to impurities Eu$_{2}$O$_{3}$ (labeled with a plus sign) and FeAs$_{2}$ (labeled with asterisks). (b) Lattice parameters $a$, $b$, $c$ and $\beta$ as functions of Ni content. (c) Crystal structure of Eu(Fe$_{1-x}$Ni$_{x})$As$_{2}$ with As zigzag chains explicitly shown. At the bottom are the Eu-ion coordinations with two sets of non-equivalent As ligands.}
\label{xrd}
\end{figure}

Figure~\ref{xrd}(a) shows the XRD patterns of the series samples of Eu(Fe$_{1-x}$Ni$_{x}$)As$_{2}$. Overall, the reflections are weak with a small signal-to-noise ratio, suggesting relatively poor crystallinity. Similar phenomena were observed in other groups~\cite{yu2017} or in other 112 materials~\cite{yakita2014,Sala2014,Yakita2015}, reflecting the inherent instability of 112 iron pnictides~\cite{kotliar.prb2017}. Nevertheless, most of the XRD reflections can be indexed with a monoclinic 112-type unit cell with $a\sim$ 3.99 {\AA}, $b\sim$ 3.90 {\AA}, $c\sim$ 10.6 {\AA}, and $\beta\sim$ 89.9$^{\circ}$~\cite{yu2017,liuyb2018}. Tiny secondary phases of Eu$_2$O$_3$ and FeAs$_2$ are discernable, yet no 122-type phase of Eu(Fe$_{1-x}$Ni$_{x}$)$_2$As$_{2}$ can be detected. The amount of each phase was quantitatively evaluated by the Rietveld analysis. The result (for details see Fig. S1 and Table S1 of the Supplemental Material (SM)~\cite{liu.SM}) indicates that most samples contain over 90 wt\% main phase, and the contents of Eu$_{2}$O$_{3}$ and FeAs$_{2}$ are only a few percent. It is emphasized that these impurities do not interfere with the magnetic properties of the samples, since no magnetism appears in either Eu$_{2}$O$_{3}$~\cite{takikawa2010.Eu2O3} or FeAs$_{2}$~\cite{yuzuri1980.FeAs2}.

The lattice constants were calculated by a least-squares fit. The obtained lattice parameters are plotted as functions of Ni content, as shown in Fig.~\ref{xrd}(b). First of all, the $\beta$ values are slightly deviated from 90$^{\circ}$, independent of Ni doping. The result indicate that Ni doping does not affect the monoclinic distortion. Secondly, the cell parameters $a$ and $b$ increase slowly with $x$. The increments are 0.0050 {\AA} (0.13\%) and 0.0053 {\AA} (0.14\%) for $a$ and $b$ axes, respectively, at $x$ = 0.15. In contrast, the $c$ axis decreases steadily, and the reduction is 0.061 {\AA} (0.58\%) $x$ = 0.15. Consequently, the cell volume (not shown in the figure) decreases by 0.31\% at $x$ = 0.15. The systematic changes in lattice parameters confirm the Ni incorporation into the lattice. The decrease in cell volume can be understood by the smaller ionic radius of Ni$^{2+}$ (compared with that of Fe$^{2+}$), whereas the slight expansion in the basal plane suggests an effective electron doping. Note that such a Ni-doping effect on the lattice constants was also observed in other iron-based pnictides such as La(Fe$_{1-x}$Ni$_{x})$AsO~\cite{cao2009}, Eu(Fe$_{1-x}$Ni$_{x}$)$_2$As$_{2}$~\cite{ren2009-1}, and RbEu(Fe$_{1-x}$Ni$_{x}$)$_4$As$_4$~\cite{ly.1144Ni}.

\subsection{\label{subsec:level2}Electrical resistivity}

\begin{figure}
\includegraphics[width=8cm]{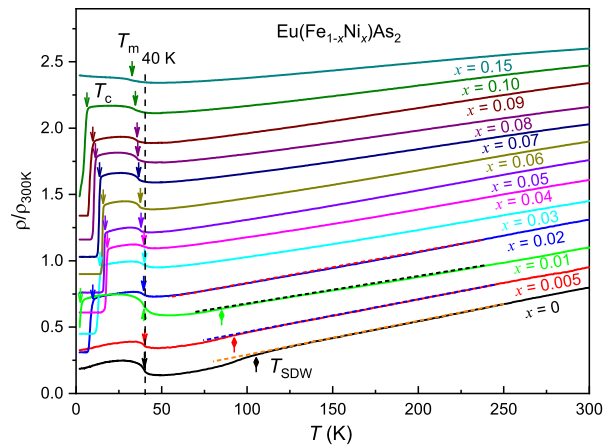}
 \caption{Temperature dependence of normalized resistivity of Eu(Fe$_{1-x}$Ni$_{x}$)As$_{2}$ polycrystalline samples. The data are shifted upward one by one for clarity. $T_\mathrm{SDW}$, $T_\mathrm{m}$, and $T_\mathrm{c}$ denote the transition temperatures of spin-density wave, Eu-spin magnetic ordering, and superconductivity, respectively. The dashed lines are a guide to the eye.}
 \label{rt}
\end{figure}

Figure~\ref{rt} shows the temperature dependence of normalized electrical resistivity, $\rho(T)/\rho_{\mathrm{300K}}$, for the Eu(Fe$_{1-x}$Ni$_{x}$)As$_{2}$ ploycrystalline samples. At high temperatures $\rho(T)$ is almost linear, while a few anomalies appear at low temperatures. As for the parent compound EuFeAs$_{2}$, a resistivity upturn at $T_\mathrm{m}\sim$ 40 K and a resistivity kink at $T_\mathrm{SDW}\sim$ 105 K can be clearly seen, which are attributed to the magnetic ordering in the Eu sublattice and the SDW transition in the Fe sublattice, respectively~\cite{yu2017,liuyb2018,albedah2020}. With a very slight Ni doping at $x$ = 0.005, the SDW anomaly is weakened, and $T_\mathrm{SDW}$ is significantly reduced. The SDW-related kink cannot be detected at $x\geq$ 0.02. Nevertheless, recent M$\mathrm{\ddot{o}}$ssbauer measurement reveals $T_\mathrm{SDW}$ = 56.2 K even at $x$ = 0.03~\cite{albedah2020}, suggesting that measurement of $\rho(T)$ is not so sensitive to the SDW ordering. Accompanied with the suppression of SDW, SC emerges at low temperatures for the samples with 0.01 $\leq x\leq$ 0.1. Remarkably, upon only 1\% Ni doping, a superconducting transition can be seen at $T_\mathrm{c}^{\mathrm{onset}}\sim$ 4 K. $T_\mathrm{c}$ first increases with $x$, achieving the maximum at $x$ = 0.04, and then decreases. Finally, no SC can be observed above 1.8 K for $x$ = 0.15. The transition temperatures of $T_\mathrm{SDW}^{\rho}$, $T_\mathrm{m}^{\rho}$, and $T_\mathrm{c}^{\rho}$, derived from the $\rho(T)$ curves, are given in Table~\ref{Tab} where $T_\mathrm{c}^{\rho}$ refers to as the resistive transition midpoint.

Unlike to the dramatic reduction of $T_\mathrm{SDW}$ and the non-monotonic changes of $T_\mathrm{c}$ associated with the electronic states in the FeAs layers, the Eu-spin ordering temperature $T_\mathrm{m}$ decreases very slowly with the Ni doping. Furthermore, no resistivity anomalies associated with the reentrant spin-glass and spin-reoriented transitions at $T_\mathrm{SG}\sim$ 15 K and $T_\mathrm{SR}\sim$ 7 K (see Section~\ref{subsec:level3} below) can be detected. It is noted that the resistivity anomaly related to the magnetic ordering at $T_\mathrm{m}$ is quite unusual. The typical response is a resistivity kink, as is seen in EuFe$_2$As$_{2}$~\cite{jiang2009} and EuNi$_{1.95}$As$_{2}$~\cite{Eu122Ni.PRB2019}, since the magnetic ordering reduces the random spin scattering. Here in Eu(Fe$_{1-x}$Ni$_{x}$)As$_{2}$, however, the As2 plane is demonstrated to be metallic~\cite{li2015.112,jiang2016.112} even with Dirac-cone like band dispersion~\cite{Dirac.APL2016}, which is likely to contribute the conductivity appreciably (conversely this also explains relatively weak anomaly at the SDW transition). The observed resistivity upturn suggests that the magnetic ordering at $T_\mathrm{m}$ probably affects the low-energy electronic state in the As2 plane. Further investigations using angle-resolved photoemission spectroscopy technique may clarify this issue.

\begin{figure}
\includegraphics[width=8cm]{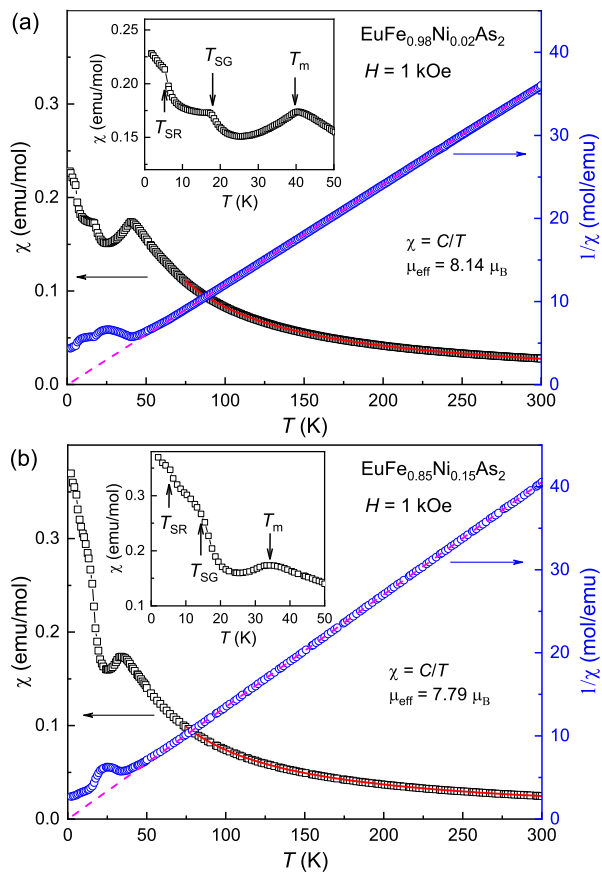}
\caption{Temperature dependence of magnetic susceptibility for Eu(Fe$_{0.98}$Ni$_{0.02})$As$_{2}$ (a) and Eu(Fe$_{0.85}$Ni$_{0.15})$As$_{2}$ (b) under a magnetic field of 1 kOe. The right axis plots the reciprocal of susceptibility. The solid lines show the fitting with Curie law in the temperature from 75 to 300 K. The insets are the close-ups which clearly show the successive magnetic transitions at $T_\mathrm{SG}$ and $T_\mathrm{SR}$ (See text for the details).}
\label{cw}
\end{figure}

\begin{table*}
\caption{Summary of physical-property parameters of Eu(Fe$_{1-x}$Ni$_{x}$)As$_{2}$ derived from the measurements of electrical resistivity (with superscript $\rho$) and magnetic susceptibility (with superscript $\chi$). $T_\mathrm{SDW}$, $T_\mathrm{c}$, $T_\mathrm{m}$, $T_\mathrm{SG}$, and $T_\mathrm{SR}$ denote the temperatures of spin-density-wave transition, superconducting transition, Eu-spin ordering, Eu-spin-glass reentrance, and Eu-spin reorientation, respectively. $\mu_\mathrm{eff}$ is the effective magnetic moment obtained from the data fitting with Curie law. The symbol ``-'' represents the case that was not able to be detected.}
\begin{ruledtabular}
\begin{tabular}{ccccccccc}
$x$ & $T_\mathrm{SDW}^{\rho}$(K)&$T_\mathrm{c}^{\rho}$(K)&$T_\mathrm{c}^{\chi}$(K)&$T_\mathrm{m}^{\rho}$(K)
&$T_\mathrm{m}^{\chi}$(K)&$T_\mathrm{SG}^{\chi}$(K)& $T_\mathrm{SR}^{\chi}$(K) &$\mu_\mathrm{eff}$($\mu_\mathrm{B}$/Eu)\\
\hline
0     &105&-&-&40.0&39.5&15.5& 5.4 &7.98 \\
0.005 &92&-&-&39.7&39.7&15.7& 5.3 &8.00 \\
0.01 &85&2.7&2.2&39.5&39.5&15.5& 5.3 &8.24 \\
0.02  &-&8.4&8.2&39.5&39.5&15.2& 5.3 &  8.14 \\
\hline
0.03  &56.2\textsuperscript{\cite{albedah2020}}&13.1&12.9&39.8&39.6&14.7& 5.1 &7.78 \\
0.04  &-&17.6&17.6&39.7&38.5&15.5& 5.2 &8.01\\
0.05  &-&16.4&16.1&37.7&37.2&16.5& 4.5 &7.74 \\
0.06  &-&15.4&15.2&37.1&36.7&15.6&-  &8.09\\
\hline
0.07  &-&13.4&13.2&36.5&36.7&15.7&- &7.81 \\
0.08  &-&10.5&10.2&36.1&35.6&15.7&5.1  &8.08 \\
0.09  &-&8.1&7.6&35.7&34.9&15.7& 5.1 &  7.98 \\
0.10   &-&4.4&-&34.7&34.2&15.6& 5.3 &  7.64 \\
0.15  &-&-&-&32.8&31.6&15.5& 5.4 &7.79 \\
\end{tabular}
\end{ruledtabular}
\label{Tab}
\end{table*}

\subsection{\label{subsec:level3}Magnetic properties}

The magnetic susceptibility ($\chi$) of Eu(Fe$_{1-x}$Ni$_{x}$)As$_{2}$ demonstrates dominant Curie-Weiss (CW) paramagnetism above the magnetic transition temperature, $T_\mathrm{m}\sim$ 40 K. Figure~\ref{cw} shows the magnetic susceptibility data of two representative samples with $x$ = 0.02 and 0.15. As can be seen, $1/\chi$ is virtually linear with $T$ at high temperatures, and its extrapolated intercept goes to zero temperature. The result indicates small values of $\chi_0$ and $\Theta$ in the general CW formula of $\chi = \chi_{0}+C/(T-\Theta)$. Indeed, such a CW fit yields $\chi_0$ fluctuating from $-7.5\times10^{-4}$ to $6.3\times10^{-4}$ emu/mol, and $\Theta$ fluctuating from $-$2.2 to 3.2 K. The scattered results are related to the mutual correlations between $\chi_0$ and $\Theta$ in the data fitting. Nevertheless, it is definite that $\Theta$ is much lower than $T_\mathrm{m}$, suggesting existence of both antiferromagnetic and ferromagnetic interactions that basically cancel out. This is different from the 122 system of Eu(Fe$_{1-x}$Ni$_{x}$)$_2$As$_{2}$~\cite{ren2009-1} and the 1144 system of RbEu(Fe$_{1-x}$Ni$_{x}$)$_4$As$_4$~\cite{ly.1144Ni}, both of which show dominant ferromagnetic interactions. We will discuss on this issue later on.

We thus fit the $\chi(T)$ data using a simplified CW formula (namely, the Curie law), $\chi = C/T$, in order to assess the effective magnetic moment reliably. The fitted Curie constants are $C$ = 8.294 and 7.395 emu K/mol, respectively, for $x=0.02$ and 0.15. Thus the effective magnetic moments $\mu_{\mathrm{eff}}=8.14$ and 7.79 $\mu_\mathrm{B}$ are yielded, which are very close to the theoretical value of $7.94$ $\mu_\mathrm{B}$ for a free Eu$^{2+}$ ion. In fact, the $\mu_{\mathrm{eff}}$ values of other samples in the Eu(Fe$_{1-x}$Ni$_{x}$)As$_{2}$ series fall within 7.94$\pm$0.3 $\mu_\mathrm{B}$ (see Table~\ref{Tab}), demonstrating the Eu$^{2+}$-ion state with spin $S$ = 7/2.

\begin{figure*}
\includegraphics[width=15cm]{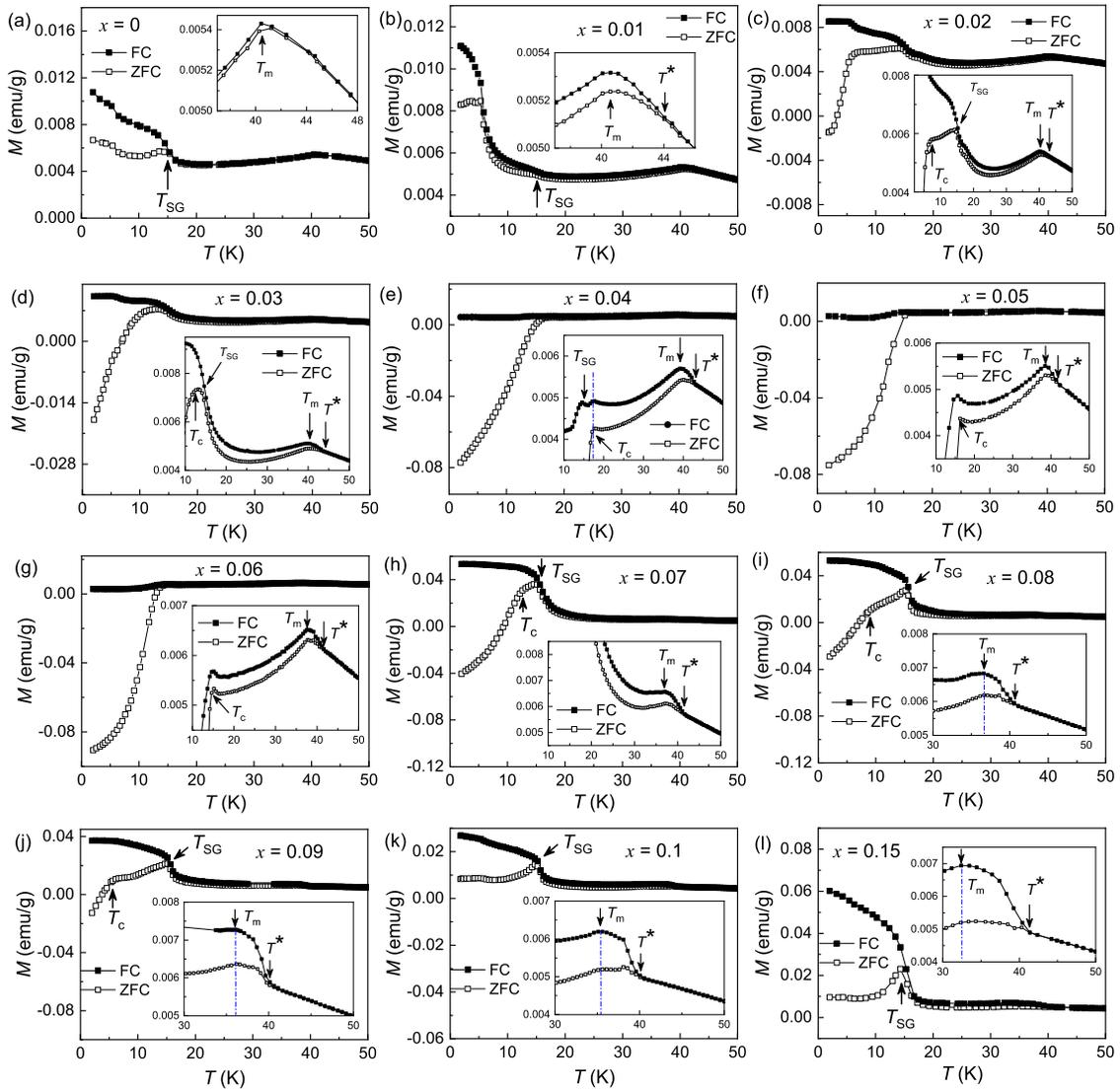}
\caption{Temperature dependence of magnetization under $H=$ 10 Oe with zero-field-cooling (ZFC) and field-cooling (FC) modes for Eu(Fe$_{1-x}$Ni$_{x}$)As$_{2}$. $T_\mathrm{c}$, $T_\mathrm{m}$, and $T_{\mathrm{SG}}$ denote the superconducting transition, Eu-spin antiferromagnetic ordering, and Eu-spin spin-glass transition, respectively. A cluster-spin-glass-like small anomaly just above $T_\mathrm{m}$ is marked with $T^{*}$.}
\label{mt}
\end{figure*}

Below $T_\mathrm{m}$, two additional anomalies at $T_\mathrm{SG}\approx$ 15 K and $T_\mathrm{SR}\approx$ 7 K can be seen in the insets of Fig.~\ref{cw}, which were respectively attributed to a spin-glass transition and a reentrant spin canting~\cite{liuyb2018}. The spin-glass transition is confirmed by the ac susceptibility measurement which shows a shift of $T_\mathrm{SG}$ to higher temperatures with increasing the frequency of the ac field applied (see Fig. S2 in the SM~\cite{liu.SM}). Meanwhile, the magnetic anomaly at $T_\mathrm{SR}$ is also observable in the ac susceptibility measurement. Considering the reentrant spin-glass transition in EuFe$_2$(As$_{1-x}$P$_{x}$)$_2$ and other related systems~\cite{zapf,dressel} and the spiral magnetic structure in EuNi$_{2}$As$_{2}$~\cite{jin2019.Eu122Ni}, we argue that the low-temperature anomalies are probably associated with a reentrance of spin glass and spin reorientation.

Figure~\ref{mt} shows the temperature dependence of magnetization under a low field of $H =$ 10 Oe in both ZFC and FC modes. The spin-glass transition is more easily identified by the bifurcations (yet the spin reorientation is less discernable). Also, the expected superconducting transition can be clearly seen with diamagnetism at low temperatures for $0.02\leq x\leq0.09$. The transition temperatures are basically consistent with those of the resistance measurements above. In the case of $T_\mathrm{c} \gtrsim T_{\mathrm{SG}}$ ($0.04\leq x\leq0.06$), the superconducting transitions can be observable in both ZFC and FC data. The phenomenon is different from that of the Eu-containing ferromagnetic superconductors in which the $\chi_{\mathrm{FC}}$ data do not show diamagnetism below $T_\mathrm{m}$ because of the spontaneous vortex ground state~\cite{jiao2017,ly.1144Ni}. The strongest superconducting diamagnetism ($-$0.09 emu/g) is found for $x=$ 0.06, equivalent to $\sim$ 60\% of the full diamagnetism, suggesting bulk SC. For the cases with $T_\mathrm{c}<T_{\mathrm{SG}}$, however, only the ZFC data could tell the superconducting transition. This is because the spin-glass state is easily magnetized by an external field, which traps superconducting vortices in the FC mode.

The close-up plots of Fig.~\ref{mt} show that the $\chi(T)$ data actually start to bifurcate at $T^*$, a few Kelvins above $T_\mathrm{m}$ for all the samples. This bifurcation behavior is reminiscent of a cluster-spin glass. The closeness between $T_\mathrm{m}$ and $T^*$ suggests that both transitions should originate from similar magnetic exchange interactions. Also noted is that only a tiny jump in the specific heat is observable at $T^*$ (see below). Therefore, this extra anomaly should come from an extrinsic origin. Based on the poor crystallinity as revealed by XRD, the cluster-glass behavior is likely to be due to the structural distortion and/or compositional inhomogeneity. As such, these defects modify the magnetic interactions, which may lead to a cluster-glassy state in stead of a long-range antiferromagnetic order. In addition, the extrinsic defects may induce local strains, such that the magnetic transition temperature is altered according to the recent high-pressure M\"{o}ssbauer study on Eu(Fe$_{1-x}$Ni$_{x}$)As$_{2}$~\cite{bi2021.Eu112Ni}.

\begin{figure*}
\includegraphics[width=15cm]{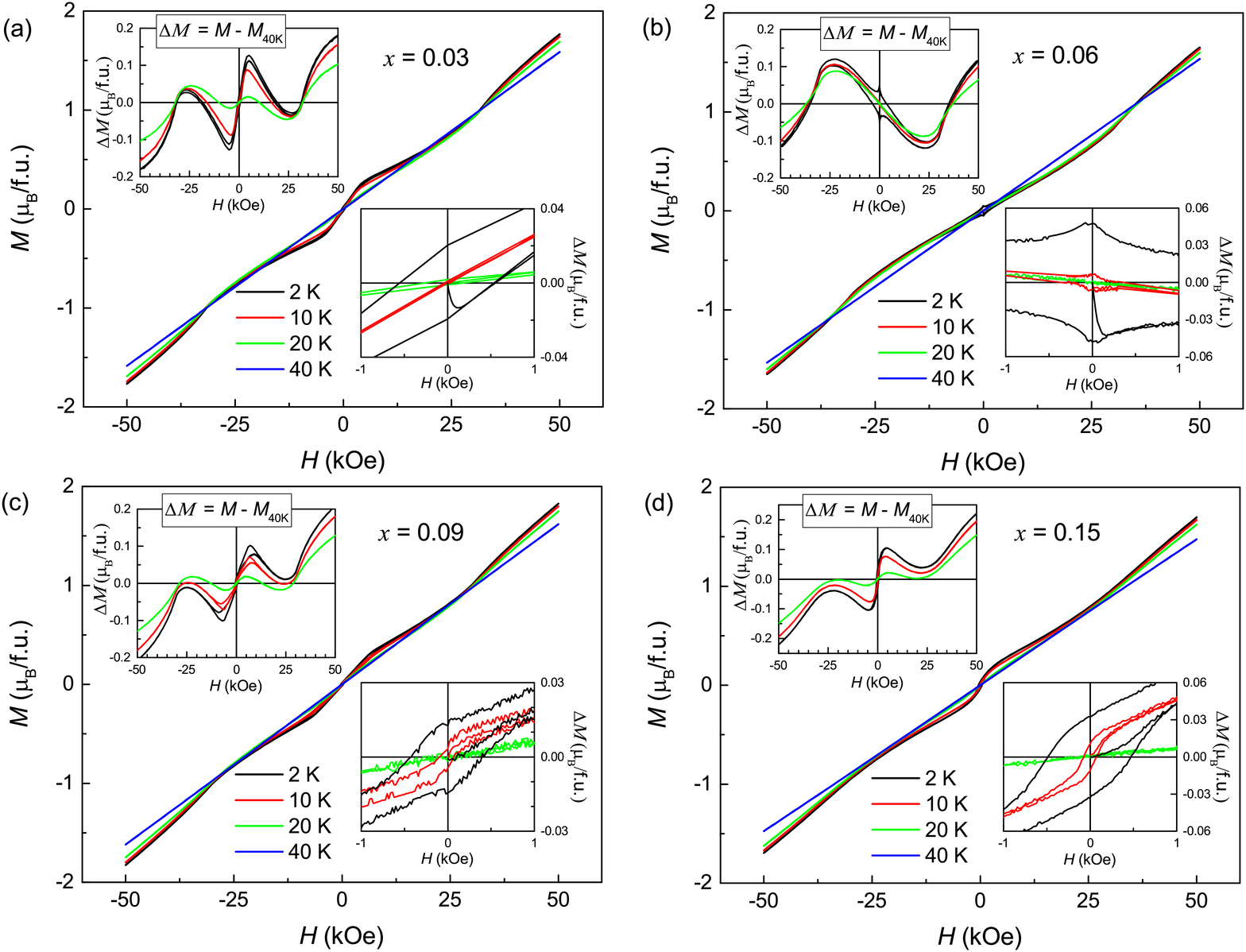}
\caption{Field dependence of magnetization at fixed temperatures for Eu(Fe$_{1-x}$Ni$_{x}$)As$_{2}$ with $x$ = 0.03 (a), $x$ = 0.06 (b), $x$ = 0.09 (c), and $x$ = 0.15 (d). The upper-left insets plot the difference in magnetization, $\Delta{M}$ = $M_{T}$ - $M_\mathrm{40K}$, and the lower-right insets show a close-up of the $\Delta{M}$ data in the low-field region.}
\label{mh}
\end{figure*}

Figure~\ref{mh} shows the field dependence of magnetization for the representative samples of Eu(Fe$_{1-x}$Ni$_{x}$)As$_{2}$ with $x$ = 0.03, 0.06, 0.09 and 0.15. The $M(H)$ curves are essentially linear at $T\geq$ 40 K (data of $T>$ 40 K are not shown for clarity). Below $T_\mathrm{m}\sim$ 40 K, $M(H)$ remains roughly linear, and the magnetization at $H=$ 50 kOe is about 1.8 $\mu_\mathrm{B}$, which is far below $gS=$ 7.0 $\mu_\mathrm{B}$ expected for ferromagnetic alignment of Eu spins. The result is in contrast with the $M(H)$ behaviors which show a saturated magnetization of $\sim$ 7.0 $\mu_\mathrm{B}$ under $H\leq$ 10 kOe in Eu(Fe$_{1-x}$Ni$_{x}$)$_2$As$_{2}$~\cite{ren2009-1}, EuFe$_{2}$(As$_{1-x}$P$_{x}$)$_{2}$~\cite{cao2011}, and RbEu(Fe$_{1-x}$Ni$_{x}$)$_4$As$_4$~\cite{ly.1144Ni}. It is thus concluded that Eu-spin antiferromagnetism is dominant in the Eu(Fe$_{1-x}$Ni$_{x}$)As$_{2}$ system.

Nevertheless, deviations from linearity are obvious in the $M(H)$ curves below 40 K. To examine the subtle changes, we made a substraction using the 40-K data as the reference, yielding $\Delta{M}=M_{T}-M_\mathrm{40K}$, as was done in our previous work~\cite{liuyb2018}. The resulted $\Delta{M}$, shown in the upper-left insets of Fig.~\ref{mh}, indicates non-monotonic variations. A metamagnetic-like transition appears at $H\approx$ 30 kOe. In the low-field regime, the $\Delta M(H)$ behavior can be basically accounted for in terms of the coexistence of SC and weak FM. In the lower-right insets of Fig.~\ref{mh}, the diamagnetism due to the superconducting transition can be observed in the initial magnetization curves for $x$ = 0.03 and 0.06. As for $x$ = 0.09, a tiny diamagnetism in the initial magnetization is still discernable, consistent with the diamagnetism shown in Fig.~\ref{mt}(j). The residual magnetization at 2 K is only about 0.015 $\mu_\mathrm{B}$/f.u., equivalent to a small internal field of $\sim$ 15 Oe. At $x$ = 0.06, a dominant superconducting contribution is observed at low temperatures, presumably because the lower critical magnetic field is higher than the spontaneous internal field (the case of Meissner state). For $x$ = 0.15, only a ferromagnetic-like hysteresis loop shows up with an enhanced residual magnetization of 0.033 $\mu_\mathrm{B}$/f.u. at 2 K.

\subsection{\label{subsec:level4}Heat capacity}

\begin{figure}
\includegraphics[width=8cm]{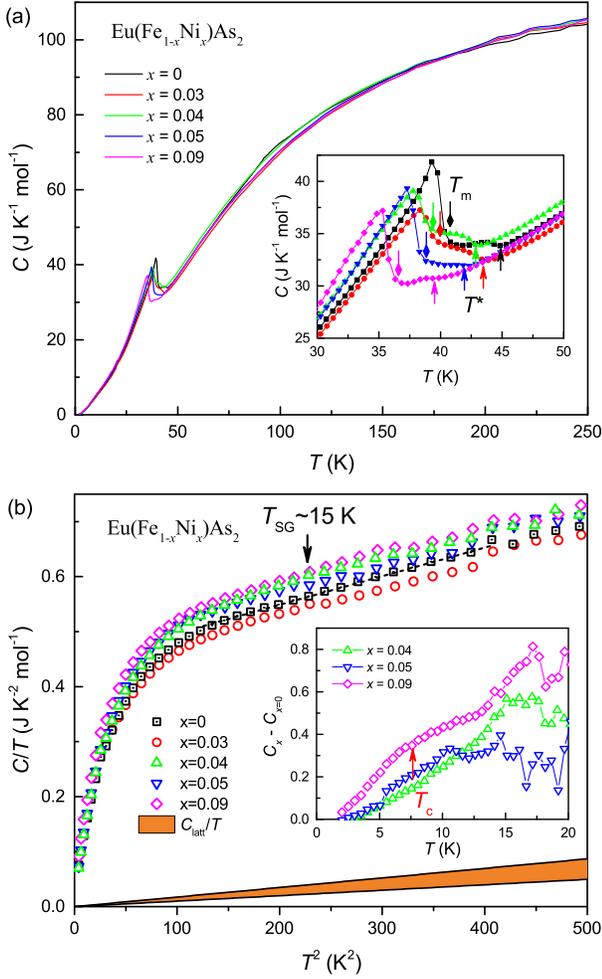}
\caption{(a) Specific heat $C$ as a function of temperature for Eu(Fe$_{1-x}$Ni$_{x}$)As$_{2}$. The inset is a close-up for showing the anomalies explicitly at $T_\mathrm{m}\sim$ 40 K. The upward and downward arrows mark the magnetic transitions at $T^{*}$ and $T_\mathrm{m}$, respectively. (b) The corresponding $C/T$ versus $T^2$ plot in the low temperatures. The position of $T_\mathrm{SG}\sim$ 15 K is marked although no obvious anomaly is discernable. The lattice contribution $C_{\mathrm{latt}}/T$ is estimated by the shaded area. The inset shows the specific-heat difference using the data of $x$ = 0 as the reference. The superconducting transition temperatures are marked with arrows.}
\label{ct}
\end{figure}

Figure~\ref{ct}(a) shows the specific-heat data for the Eu(Fe$_{1-x}$Ni$_{x}$)As$_{2}$ samples with $x$ = 0, 0.03, 0.04, 0.05 and 0.09. The Eu-spin ordering is confirmed by the specific-heat jump at $T_\mathrm{m}$, as is clearly seen in the inset. $T_\mathrm{m}$ decreases with increasing $x$, consistent with the resistive and magnetic measurement results above. Meanwhile, an additional anomaly at $T^*$ can be detected, coinciding with the magnetic susceptibility bifurcation shown in the insets of Fig.~\ref{mt}.

The low-temperature specific-heat data were analyzed in terms of $C/T$ as a function of $T^2$, which is shown in Fig.~\ref{ct}(b). The lattice contribution was estimated using Debye law, $C_L=\beta T^3$, where $\beta=(12/5)NR\pi^{4}/\Theta_D^3$. The Debye temperatures from 355 K to 428 K are taken from the M\"{o}ssbauer measurement~\cite{albedah2020}. The resulted lattice contribution is found to be much lower than the magnetic contributions, indicating dominant contributions from the Eu-spin magnetism. Note that no obvious anomalies can be seen in the successive magnetic transitions at $T_{\mathrm{SG}}$ and $T_{\mathrm{SR}}$, indicating no entropy changes at the reentrant transitions. Shown in the inset of Fig.~\ref{ct}(b) is the specific-heat difference by subtracting the $C(T)$ of $x$ = 0. The data above 15 K is noisy due to some unknown reasons, which makes it difficult to identify the superconducting transitions for $x$ = 0.04 and 0.05. In the case of $x$ = 0.09, the slope change at around 8 K could be due to the superconducting transition.

\subsection{\label{subsec:level5}Phase diagram and Discussion}

\begin{figure}
\includegraphics[width=8cm]{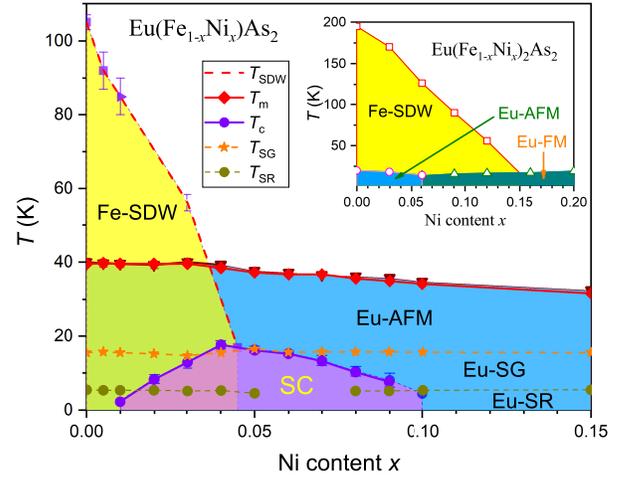}
\caption{(a) Magnetic and superconducting phase diagram of Eu(Fe$_{1-x}$Ni$_{x}$)As$_{2}$. $T_\mathrm{SDW}$, $T_\mathrm{m}$, $T_\mathrm{c}$, $T_\mathrm{SG}$, and $T_\mathrm{SR}$ denote the temperatures of spin-density wave (SDW) ordering, Eu-spin antiferromagnetic (AFM) ordering, superconducting transition, spin-glass (SG) reentrance, and spin reorientation (SR), respectively. For comparison, the inset shows the phase diagram of Eu(Fe$_{1-x}$Ni$_{x}$)$_{2}$As$_{2}$ (adapted from Ref.~\cite{ren2009-1}).}
\label{pd}
\end{figure}

Using the data summarized in Table~\ref{Tab}, we mapped out the magnetic and superconducting phase diagram of Eu(Fe$_{1-x}$Ni$_{x}$)As$_{2}$, which is shown in Fig.~\ref{pd}. Firstly, the SDW order is suppressed with the Ni doping, and it survives up to $x\sim$ 0.04. Meanwhile SC emerges from $x=$ 0.01 to $x=$ 0.1. As a result, there is an overlap ($0.01\leq x\leq0.04$) exhibiting coexistence of SC and SDW. Coincidentally, the maximal $T_\mathrm{c}$ shows up at which the SDW tends to disappear. Secondly, the antiferromagnetic transition temperature $T_\mathrm{m}$ associated with the Eu spins decreases steadily and slowly with the Ni doping. The decrease of $T_\mathrm{m}$ may be associated with the the electron doping as well as the slight increases of $a$ and $b$ axes as shown above. The Eu-spin ordered state undergoes reentrant spin-glass and spin-reorientation transitions at $T_\mathrm{SG}\sim$ 15 K and $T_\mathrm{SR}\sim$ 7 K, respectively, independent of Ni doping. The $T_\mathrm{SG}$ and $T_\mathrm{SR}$ lines cut the superconducting dome on the top and at the bottom, respectively, which gives rise to rich phenomena of the interplay between SC and weak FM.

The phase diagram associated with FeAs layers is commonly seen in other iron-based superconductors~\cite{alberto2016,luo2015}. It is well known that Ni doping in iron-pnictide systems introduces extra itinerant electrons, which generally suppresses the SDW order and then leads to SC~\cite{li2009,cao2009}. Therefore, the present system gives an additional example demonstrating the universal superconducting phase diagram. Nevertheless, the details differ. One is that SC appears at a slight Ni-doping level, yet with an SDW ordering temperature as high as 85 K. The second is that the optimal Ni-doping level (with a maximum $T_\mathrm{c}$) is $x=$ 0.04, which is less than half of that in 122-type Ba(Fe$_{1-x}$Ni$_{x}$)$_{2}$As$_{2}$~\cite{li2009}. These two characters suggest that the 112 system is closer to a superconducting state. Thirdly, the SC region is much broader than that of La(Fe$_{1-x}$Ni$_{x}$)AsO~\cite{cao2009}, suggesting that the doping-induced disorder is much less significant in the 112 system. Indeed, there have been many reports in 112 systems showing the effectiveness of Fe-site dopings for achieving bulk SC~\cite{xing2015.112Co,jiang2016.112,xie2017}. 

It is informative to compare the phase diagram with that of the 122-type Eu(Fe$_{1-x}$Ni$_{x}$)$_{2}$As$_{2}$. As is shown in the inset of Fig.~\ref{pd}, the most striking difference is that SC is absent in Eu(Fe$_{1-x}$Ni$_{x}$)$_{2}$As$_{2}$. In most Eu-containing 122 systems, as a matter of fact, SC is overwhelmed by the Eu-spin magnetism when $T_\mathrm{m}>T_\mathrm{c}$~\cite{cao2011,dressel}. By contrast, SC is robust for $T_\mathrm{m}\sim $ 40 K in the present 112 system. Additionally, the maximal $T_\mathrm{c}$ (17.6 K) is close to the counterpart (20.5 K) of an Eu-free system of Ba(Fe$_{1-x}$Ni$_{x}$)$_{2}$As$_{2}$~\cite{li2009}, suggesting an insignificant influence of Eu spins on SC. Another obvious difference is about the Eu-spin ordering. In the 122 system, the Eu-spin ordering temperature is about 18 K, and the Eu spins become ferromagnetically ordered at the high Ni-doping regime~\cite{ren2009-1}. In the present 112 system, however, $T_\mathrm{m}$ is much higher and, the $T_\mathrm{SG}$ is close, to the Eu-spin ordering temperature of the 122 system.

Then, how can we understand the complex Eu-spin magnetism? The crystal structure shown in Fig.~\ref{xrd}(c) gives a clue to the  answer. Unlike the symmetrical coordinations of Eu$^{2+}$ ions in the 122 system, the Eu$^{2+}$ ions in Eu(Fe$_{1-x}$Ni$_{x}$)As$_{2}$ are asymmetrically coordinated by four As2 atoms in the zigzag-As plane and another four As1 atoms from FeAs layers. The Eu-spin ordering temperatures of EuFe$_2$As$_{2}$ and EuNi$_2$As$_{2}$ are 19 K and 14 K, respectively~\cite{Raffius1993}. Therefore, the RKKY interactions mediated by the conduction electrons in (Fe,Ni)As layers are not likely to account for the 40-K magnetic transition. Instead, the metallic As2 sheets probably play the dominant role for the 40-K antiferromagnetism (this naturally explains the robustness of SC in the presence of the antiferromagnetism). On the other hand, the RKKY interactions mediated by the FeAs layers is basically ferromagnetic~\cite{jiang2009}. Then, there must be a competition between AFM and FM within an Eu layer, which could cause the observed spin glass and spin reorientation. A similar competition between AFM and FM was proposed for explanation of the spin-glass reentrance in EuFe$_2$(As,P)$_{2}$~\cite{zapf}. Note that the dominant ferromagnetic interactions mediated by FeAs layers tend to cancels out the antiferromagnetic interactions mediated by the As sheets in EuFeAs$_{2}$. Thus the vanishing small value of CW temperature (in spite of a high N\'{e}el temperature) in EuFeAs$_{2}$ can be understood.


\section{\label{sec:level4}Concluding remarks}

We have successfully synthesized nearly single-phase samples of 112-type iron pnictide Eu(Fe$_{1-x}$Ni$_{x}$)As$_{2}$ ($0\leq x\leq 0.15$). The gradual incorporation of Ni is corroborated by the systematic changes in lattice parameters. SC emerges for a slight Ni doping with $x=$ 0.01. The superconducting transition temperature $T_\mathrm{c}$ first increases and then decreases, showing a maximum of 17.6 K at $x$ = 0.04. Meanwhile, the SDW order is suppressed by the Ni doping, and it disappears at $x\geq$ 0.05. The superconducting phase diagram is qualitatively similar to those of most iron-based superconductors.

The Eu spins have little impact on the SDW and SC, meanwhile, the complex Eu-spin magnetism is only slightly influenced by the Ni doping. The dominant antiferromagnetic ordering at $T_\mathrm{m}\sim$ 40 K is probably due to the RKKY interactions mediated by the metallic spacer layers composed of zigzag As chains, and the subsequent spin-glass and spin-reorientation reentrances are associated with the competition between the AFM of Eu (mediated by the As2 atoms) and the FM interactions mediated by the FeAs layers. The robustness of SC against the exchange interactions suggests that the Fe-3d conduction electrons selectively participate in the superconducting Cooper pairing (the conduction electrons involved in the RKKY interaction do not contribute to SC)~\cite{cao2011}. Finally, with the zigzag As chains and the reentrant Eu-spin magnetism, one expects that the Eu(Fe$_{1-x}$Ni$_{x}$)As$_{2}$ system could exhibit additionally interesting topological states as theories predicted~\cite{wuxx.prb2015,wuxx.PRX2020}.\\

\begin{acknowledgments}
This work was supported by the National Key Research and Development Program of China (2016YFA0300202), the National Natural Science Foundation of China (12004337), the Key R \& D Program of Zhejiang Province, China (2021C01002), and the Fundamental Research Funds for the Central Universities of China.
\end{acknowledgments}

\section*{References}

\bibliographystyle{apsrev4-2}

\begin{thebibliography}{51}%
\makeatletter
\providecommand \@ifxundefined [1]{%
 \@ifx{#1\undefined}
}%
\providecommand \@ifnum [1]{%
 \ifnum #1\expandafter \@firstoftwo
 \else \expandafter \@secondoftwo
 \fi
}%
\providecommand \@ifx [1]{%
 \ifx #1\expandafter \@firstoftwo
 \else \expandafter \@secondoftwo
 \fi
}%
\providecommand \natexlab [1]{#1}%
\providecommand \enquote  [1]{``#1''}%
\providecommand \bibnamefont  [1]{#1}%
\providecommand \bibfnamefont [1]{#1}%
\providecommand \citenamefont [1]{#1}%
\providecommand \href@noop [0]{\@secondoftwo}%
\providecommand \href [0]{\begingroup \@sanitize@url \@href}%
\providecommand \@href[1]{\@@startlink{#1}\@@href}%
\providecommand \@@href[1]{\endgroup#1\@@endlink}%
\providecommand \@sanitize@url [0]{\catcode `\\12\catcode `\$12\catcode
  `\&12\catcode `\#12\catcode `\^12\catcode `\_12\catcode `\%12\relax}%
\providecommand \@@startlink[1]{}%
\providecommand \@@endlink[0]{}%
\providecommand \url  [0]{\begingroup\@sanitize@url \@url }%
\providecommand \@url [1]{\endgroup\@href {#1}{\urlprefix }}%
\providecommand \urlprefix  [0]{URL }%
\providecommand \Eprint [0]{\href }%
\providecommand \doibase [0]{https://doi.org/}%
\providecommand \selectlanguage [0]{\@gobble}%
\providecommand \bibinfo  [0]{\@secondoftwo}%
\providecommand \bibfield  [0]{\@secondoftwo}%
\providecommand \translation [1]{[#1]}%
\providecommand \BibitemOpen [0]{}%
\providecommand \bibitemStop [0]{}%
\providecommand \bibitemNoStop [0]{.\EOS\space}%
\providecommand \EOS [0]{\spacefactor3000\relax}%
\providecommand \BibitemShut  [1]{\csname bibitem#1\endcsname}%
\let\auto@bib@innerbib\@empty
\bibitem [{\citenamefont {Cao}\ \emph {et~al.}(2012)\citenamefont {Cao},
  \citenamefont {Jiao}, \citenamefont {Luo}, \citenamefont {Ren}, \citenamefont
  {Jiang},\ and\ \citenamefont {Xu}}]{cao2012}%
  \BibitemOpen
  \bibfield  {author} {\bibinfo {author} {\bibfnamefont {G.-H.}\ \bibnamefont
  {Cao}}, \bibinfo {author} {\bibfnamefont {W.-H.}\ \bibnamefont {Jiao}},
  \bibinfo {author} {\bibfnamefont {Y.-K.}\ \bibnamefont {Luo}}, \bibinfo
  {author} {\bibfnamefont {Z.}~\bibnamefont {Ren}}, \bibinfo {author}
  {\bibfnamefont {S.}~\bibnamefont {Jiang}},\ and\ \bibinfo {author}
  {\bibfnamefont {Z.-A.}\ \bibnamefont {Xu}},\ }\bibinfo {title} {Coexistence
  of superconductivity and ferromagnetism in iron pnictides},\ \href
  {http://stacks.iop.org/1742-6596/391/i=1/a=012123} {\bibfield  {journal}
  {\bibinfo  {journal} {J. Phys.: Conf. Ser.}\ }\textbf {\bibinfo {volume}
  {391}},\ \bibinfo {pages} {012123} (\bibinfo {year} {2012})}\BibitemShut
  {NoStop}%
\bibitem [{\citenamefont {Zapf}\ and\ \citenamefont {Dressel}(2017)}]{dressel}%
  \BibitemOpen
  \bibfield  {author} {\bibinfo {author} {\bibfnamefont {S.}~\bibnamefont
  {Zapf}}\ and\ \bibinfo {author} {\bibfnamefont {M.}~\bibnamefont {Dressel}},\
  }\bibinfo {title} {Europium-based iron pnictides: a unique laboratory for
  magnetism, superconductivity and structural effects},\ \href
  {http://stacks.iop.org/0034-4885/80/i=1/a=016501} {\bibfield  {journal}
  {\bibinfo  {journal} {Rep. Prog. Phys.}\ }\textbf {\bibinfo {volume} {80}},\
  \bibinfo {pages} {016501} (\bibinfo {year} {2017})}\BibitemShut {NoStop}%
\bibitem [{\citenamefont {Ren}\ \emph {et~al.}(2009{\natexlab{a}})\citenamefont
  {Ren}, \citenamefont {Tao}, \citenamefont {Jiang}, \citenamefont {Feng},
  \citenamefont {Wang}, \citenamefont {Dai}, \citenamefont {Cao},\ and\
  \citenamefont {Xu}}]{ren2009}%
  \BibitemOpen
  \bibfield  {author} {\bibinfo {author} {\bibfnamefont {Z.}~\bibnamefont
  {Ren}}, \bibinfo {author} {\bibfnamefont {Q.}~\bibnamefont {Tao}}, \bibinfo
  {author} {\bibfnamefont {S.}~\bibnamefont {Jiang}}, \bibinfo {author}
  {\bibfnamefont {C.}~\bibnamefont {Feng}}, \bibinfo {author} {\bibfnamefont
  {C.}~\bibnamefont {Wang}}, \bibinfo {author} {\bibfnamefont {J.}~\bibnamefont
  {Dai}}, \bibinfo {author} {\bibfnamefont {G.}~\bibnamefont {Cao}},\ and\
  \bibinfo {author} {\bibfnamefont {Z.}~\bibnamefont {Xu}},\ }\bibinfo {title}
  {Superconductivity Induced by Phosphorus Doping and Its Coexistence with
  Ferromagnetism in
  {${\mathrm{EuFe}}_{2}({\mathrm{As}}_{0.7}{\mathrm{P}}_{0.3}{)}_{2}$}},\ \href
  {https://doi.org/10.1103/PhysRevLett.102.137002} {\bibfield  {journal}
  {\bibinfo  {journal} {Phys. Rev. Lett.}\ }\textbf {\bibinfo {volume} {102}},\
  \bibinfo {pages} {137002} (\bibinfo {year} {2009}{\natexlab{a}})}\BibitemShut
  {NoStop}%
\bibitem [{\citenamefont {Cao}\ \emph {et~al.}(2011)\citenamefont {Cao},
  \citenamefont {Xu}, \citenamefont {Ren}, \citenamefont {Jiang}, \citenamefont
  {Feng},\ and\ \citenamefont {Xu}}]{cao2011}%
  \BibitemOpen
  \bibfield  {author} {\bibinfo {author} {\bibfnamefont {G.}~\bibnamefont
  {Cao}}, \bibinfo {author} {\bibfnamefont {S.}~\bibnamefont {Xu}}, \bibinfo
  {author} {\bibfnamefont {Z.}~\bibnamefont {Ren}}, \bibinfo {author}
  {\bibfnamefont {S.}~\bibnamefont {Jiang}}, \bibinfo {author} {\bibfnamefont
  {C.}~\bibnamefont {Feng}},\ and\ \bibinfo {author} {\bibfnamefont
  {Z.}~\bibnamefont {Xu}},\ }\bibinfo {title} {Superconductivity and
  ferromagnetism in EuFe$_{2}$(As$_{1-x}$P$_{x}$)$_{2}$},\ \href
  {https://doi.org/10.1088/0953-8984/23/46/464204} {\bibfield  {journal}
  {\bibinfo  {journal} {J. Phys.: Condens. Matt.}\ }\textbf {\bibinfo {volume}
  {23}},\ \bibinfo {pages} {464204} (\bibinfo {year} {2011})}\BibitemShut
  {NoStop}%
\bibitem [{\citenamefont {Pogrebna}\ \emph {et~al.}(2015)\citenamefont
  {Pogrebna}, \citenamefont {Mertelj}, \citenamefont {Vuji\u{c}i\'{c}},
  \citenamefont {Cao}, \citenamefont {Xu},\ and\ \citenamefont
  {Mihailovic}}]{Pogrebna2015}%
  \BibitemOpen
  \bibfield  {author} {\bibinfo {author} {\bibfnamefont {A.}~\bibnamefont
  {Pogrebna}}, \bibinfo {author} {\bibfnamefont {T.}~\bibnamefont {Mertelj}},
  \bibinfo {author} {\bibfnamefont {N.}~\bibnamefont {Vuji\u{c}i\'{c}}},
  \bibinfo {author} {\bibfnamefont {G.}~\bibnamefont {Cao}}, \bibinfo {author}
  {\bibfnamefont {Z.~A.}\ \bibnamefont {Xu}},\ and\ \bibinfo {author}
  {\bibfnamefont {D.}~\bibnamefont {Mihailovic}},\ }\bibinfo {title}
  {Coexistence of ferromagnetism and superconductivity in iron based pnictides:
  a time resolved magnetooptical study},\ \href
  {https://doi.org/10.1038/srep07754} {\bibfield  {journal} {\bibinfo
  {journal} {Sci. Rep.}\ }\textbf {\bibinfo {volume} {5}},\ \bibinfo {pages}
  {7754} (\bibinfo {year} {2015})}\BibitemShut {NoStop}%
\bibitem [{\citenamefont {Nowik}\ \emph {et~al.}(2011)\citenamefont {Nowik},
  \citenamefont {Felner}, \citenamefont {Ren}, \citenamefont {Cao},\ and\
  \citenamefont {Xu}}]{felner2011}%
  \BibitemOpen
  \bibfield  {author} {\bibinfo {author} {\bibfnamefont {I.}~\bibnamefont
  {Nowik}}, \bibinfo {author} {\bibfnamefont {I.}~\bibnamefont {Felner}},
  \bibinfo {author} {\bibfnamefont {Z.}~\bibnamefont {Ren}}, \bibinfo {author}
  {\bibfnamefont {G.~H.}\ \bibnamefont {Cao}},\ and\ \bibinfo {author}
  {\bibfnamefont {Z.~A.}\ \bibnamefont {Xu}},\ }\bibinfo {title} {Coexistence
  of ferromagnetism and superconductivity: magnetization and M\"{o}ssbauer
  studies of {EuFe$_{2}$(As$_{1-x}$P$_{x}$)$_{2}$}},\ \href
  {http://stacks.iop.org/0953-8984/23/i=6/a=065701} {\bibfield  {journal}
  {\bibinfo  {journal} {J. Phys.: Condens. Matt.}\ }\textbf {\bibinfo {volume}
  {23}},\ \bibinfo {pages} {065701} (\bibinfo {year} {2011})}\BibitemShut
  {NoStop}%
\bibitem [{\citenamefont {Nandi}\ \emph
  {et~al.}(2014{\natexlab{a}})\citenamefont {Nandi}, \citenamefont {Jin},
  \citenamefont {Xiao}, \citenamefont {Su}, \citenamefont {Price},
  \citenamefont {Schmidt}, \citenamefont {Schmalzl}, \citenamefont {Chatterji},
  \citenamefont {Jeevan}, \citenamefont {Gegenwart},\ and\ \citenamefont
  {Br\"uckel}}]{nd2014}%
  \BibitemOpen
  \bibfield  {author} {\bibinfo {author} {\bibfnamefont {S.}~\bibnamefont
  {Nandi}}, \bibinfo {author} {\bibfnamefont {W.~T.}\ \bibnamefont {Jin}},
  \bibinfo {author} {\bibfnamefont {Y.}~\bibnamefont {Xiao}}, \bibinfo {author}
  {\bibfnamefont {Y.}~\bibnamefont {Su}}, \bibinfo {author} {\bibfnamefont
  {S.}~\bibnamefont {Price}}, \bibinfo {author} {\bibfnamefont
  {W.}~\bibnamefont {Schmidt}}, \bibinfo {author} {\bibfnamefont
  {K.}~\bibnamefont {Schmalzl}}, \bibinfo {author} {\bibfnamefont
  {T.}~\bibnamefont {Chatterji}}, \bibinfo {author} {\bibfnamefont {H.~S.}\
  \bibnamefont {Jeevan}}, \bibinfo {author} {\bibfnamefont {P.}~\bibnamefont
  {Gegenwart}},\ and\ \bibinfo {author} {\bibfnamefont {T.}~\bibnamefont
  {Br\"uckel}},\ }\bibinfo {title} {Coexistence of ferromagnetism and
  superconductivity in iron based pnictides: a time resolved magnetooptical
  study},\ \href {https://doi.org/10.1103/PhysRevB.90.094407} {\bibfield
  {journal} {\bibinfo  {journal} {Phys. Rev. B}\ }\textbf {\bibinfo {volume}
  {90}},\ \bibinfo {pages} {094407} (\bibinfo {year}
  {2014}{\natexlab{a}})}\BibitemShut {NoStop}%
\bibitem [{\citenamefont {Nandi}\ \emph
  {et~al.}(2014{\natexlab{b}})\citenamefont {Nandi}, \citenamefont {Jin},
  \citenamefont {Xiao}, \citenamefont {Su}, \citenamefont {Price},
  \citenamefont {Shukla}, \citenamefont {Strempfer}, \citenamefont {Jeevan},
  \citenamefont {Gegenwart},\ and\ \citenamefont {Br\"uckel}}]{rxs2014}%
  \BibitemOpen
  \bibfield  {author} {\bibinfo {author} {\bibfnamefont {S.}~\bibnamefont
  {Nandi}}, \bibinfo {author} {\bibfnamefont {W.~T.}\ \bibnamefont {Jin}},
  \bibinfo {author} {\bibfnamefont {Y.}~\bibnamefont {Xiao}}, \bibinfo {author}
  {\bibfnamefont {Y.}~\bibnamefont {Su}}, \bibinfo {author} {\bibfnamefont
  {S.}~\bibnamefont {Price}}, \bibinfo {author} {\bibfnamefont {D.~K.}\
  \bibnamefont {Shukla}}, \bibinfo {author} {\bibfnamefont {J.}~\bibnamefont
  {Strempfer}}, \bibinfo {author} {\bibfnamefont {H.~S.}\ \bibnamefont
  {Jeevan}}, \bibinfo {author} {\bibfnamefont {P.}~\bibnamefont {Gegenwart}},\
  and\ \bibinfo {author} {\bibfnamefont {T.}~\bibnamefont {Br\"uckel}},\
  }\bibinfo {title} {Coexistence of superconductivity and ferromagnetism in
  {P}-doped {${\mathrm{EuFe}}_{2}{\mathrm{As}}_{2}$}},\ \href
  {https://doi.org/10.1103/PhysRevB.89.014512} {\bibfield  {journal} {\bibinfo
  {journal} {Phys. Rev. B}\ }\textbf {\bibinfo {volume} {89}},\ \bibinfo
  {pages} {014512} (\bibinfo {year} {2014}{\natexlab{b}})}\BibitemShut
  {NoStop}%
\bibitem [{\citenamefont {Zapf}\ \emph {et~al.}(2013)\citenamefont {Zapf},
  \citenamefont {Jeevan}, \citenamefont {Ivek}, \citenamefont {Pfister},
  \citenamefont {Klingert}, \citenamefont {Jiang}, \citenamefont {Wu},
  \citenamefont {Gegenwart}, \citenamefont {Kremer},\ and\ \citenamefont
  {Dressel}}]{zapf}%
  \BibitemOpen
  \bibfield  {author} {\bibinfo {author} {\bibfnamefont {S.}~\bibnamefont
  {Zapf}}, \bibinfo {author} {\bibfnamefont {H.~S.}\ \bibnamefont {Jeevan}},
  \bibinfo {author} {\bibfnamefont {T.}~\bibnamefont {Ivek}}, \bibinfo {author}
  {\bibfnamefont {F.}~\bibnamefont {Pfister}}, \bibinfo {author} {\bibfnamefont
  {F.}~\bibnamefont {Klingert}}, \bibinfo {author} {\bibfnamefont
  {S.}~\bibnamefont {Jiang}}, \bibinfo {author} {\bibfnamefont
  {D.}~\bibnamefont {Wu}}, \bibinfo {author} {\bibfnamefont {P.}~\bibnamefont
  {Gegenwart}}, \bibinfo {author} {\bibfnamefont {R.~K.}\ \bibnamefont
  {Kremer}},\ and\ \bibinfo {author} {\bibfnamefont {M.}~\bibnamefont
  {Dressel}},\ }\bibinfo {title}
  {{${\mathrm{EuFe}}_{2}({\mathrm{As}}_{1-x}{\mathrm{P}}_{x}{)}_{2}$}:
  Reentrant Spin Glass and Superconductivity},\ \href
  {https://doi.org/10.1103/PhysRevLett.110.237002} {\bibfield  {journal}
  {\bibinfo  {journal} {Phys. Rev. Lett.}\ }\textbf {\bibinfo {volume} {110}},\
  \bibinfo {pages} {237002} (\bibinfo {year} {2013})}\BibitemShut {NoStop}%
\bibitem [{\citenamefont {Kawashima}\ \emph {et~al.}(2016)\citenamefont
  {Kawashima}, \citenamefont {Kinjo}, \citenamefont {Nishio}, \citenamefont
  {Ishida}, \citenamefont {Fujihisa}, \citenamefont {Gotoh}, \citenamefont
  {Kihou}, \citenamefont {Eisaki}, \citenamefont {Yoshida},\ and\ \citenamefont
  {Iyo}}]{Eu1144}%
  \BibitemOpen
  \bibfield  {author} {\bibinfo {author} {\bibfnamefont {K.}~\bibnamefont
  {Kawashima}}, \bibinfo {author} {\bibfnamefont {T.}~\bibnamefont {Kinjo}},
  \bibinfo {author} {\bibfnamefont {T.}~\bibnamefont {Nishio}}, \bibinfo
  {author} {\bibfnamefont {S.}~\bibnamefont {Ishida}}, \bibinfo {author}
  {\bibfnamefont {H.}~\bibnamefont {Fujihisa}}, \bibinfo {author}
  {\bibfnamefont {Y.}~\bibnamefont {Gotoh}}, \bibinfo {author} {\bibfnamefont
  {K.}~\bibnamefont {Kihou}}, \bibinfo {author} {\bibfnamefont
  {H.}~\bibnamefont {Eisaki}}, \bibinfo {author} {\bibfnamefont
  {Y.}~\bibnamefont {Yoshida}},\ and\ \bibinfo {author} {\bibfnamefont
  {A.}~\bibnamefont {Iyo}},\ }\bibinfo {title} {Superconductivity in Fe-based
  compound Eu$A$Fe$_{4}{\mathrm{As}}_{4}$ ($A$ = Rb and Cs)},\ \href@noop {}
  {\bibfield  {journal} {\bibinfo  {journal} {J. Phys. Soc. Jpn.}\ }\textbf
  {\bibinfo {volume} {85}},\ \bibinfo {pages} {064710} (\bibinfo {year}
  {2016})}\BibitemShut {NoStop}%
\bibitem [{\citenamefont {Liu}\ \emph {et~al.}(2016{\natexlab{a}})\citenamefont
  {Liu}, \citenamefont {Liu}, \citenamefont {Tang}, \citenamefont {Jiang},
  \citenamefont {Wang}, \citenamefont {Ablimit}, \citenamefont {Jiao},
  \citenamefont {Tao}, \citenamefont {Feng}, \citenamefont {Xu},\ and\
  \citenamefont {Cao}}]{liuy2016-1}%
  \BibitemOpen
  \bibfield  {author} {\bibinfo {author} {\bibfnamefont {Y.}~\bibnamefont
  {Liu}}, \bibinfo {author} {\bibfnamefont {Y.-B.}\ \bibnamefont {Liu}},
  \bibinfo {author} {\bibfnamefont {Z.-T.}\ \bibnamefont {Tang}}, \bibinfo
  {author} {\bibfnamefont {H.}~\bibnamefont {Jiang}}, \bibinfo {author}
  {\bibfnamefont {Z.-C.}\ \bibnamefont {Wang}}, \bibinfo {author}
  {\bibfnamefont {A.}~\bibnamefont {Ablimit}}, \bibinfo {author} {\bibfnamefont
  {W.-H.}\ \bibnamefont {Jiao}}, \bibinfo {author} {\bibfnamefont
  {Q.}~\bibnamefont {Tao}}, \bibinfo {author} {\bibfnamefont {C.-M.}\
  \bibnamefont {Feng}}, \bibinfo {author} {\bibfnamefont {Z.-A.}\ \bibnamefont
  {Xu}},\ and\ \bibinfo {author} {\bibfnamefont {G.-H.}\ \bibnamefont {Cao}},\
  }\bibinfo {title} {Superconductivity and ferromagnetism in hole-doped
  ${\mathrm{RbEuFe}}_{4}{\mathrm{As}}_{4}$},\ \href
  {https://doi.org/10.1103/PhysRevB.93.214503} {\bibfield  {journal} {\bibinfo
  {journal} {Phys. Rev. B}\ }\textbf {\bibinfo {volume} {93}},\ \bibinfo
  {pages} {214503} (\bibinfo {year} {2016}{\natexlab{a}})}\BibitemShut
  {NoStop}%
\bibitem [{\citenamefont {Liu}\ \emph {et~al.}(2016{\natexlab{b}})\citenamefont
  {Liu}, \citenamefont {Liu}, \citenamefont {Chen}, \citenamefont {Tang},
  \citenamefont {Jiao}, \citenamefont {Tao}, \citenamefont {Xu},\ and\
  \citenamefont {Cao}}]{liuy2016-2}%
  \BibitemOpen
  \bibfield  {author} {\bibinfo {author} {\bibfnamefont {Y.}~\bibnamefont
  {Liu}}, \bibinfo {author} {\bibfnamefont {Y.-B.}\ \bibnamefont {Liu}},
  \bibinfo {author} {\bibfnamefont {Q.}~\bibnamefont {Chen}}, \bibinfo {author}
  {\bibfnamefont {Z.-T.}\ \bibnamefont {Tang}}, \bibinfo {author}
  {\bibfnamefont {W.-H.}\ \bibnamefont {Jiao}}, \bibinfo {author}
  {\bibfnamefont {Q.}~\bibnamefont {Tao}}, \bibinfo {author} {\bibfnamefont
  {Z.-A.}\ \bibnamefont {Xu}},\ and\ \bibinfo {author} {\bibfnamefont {G.-H.}\
  \bibnamefont {Cao}},\ }\bibinfo {title} {A new ferromagnetic superconductor:
  CsEuFe$_{4}$As$_{4}$},\ \href@noop {} {\bibfield  {journal} {\bibinfo
  {journal} {Sci. Bull.}\ }\textbf {\bibinfo {volume} {61}},\ \bibinfo {pages}
  {1213} (\bibinfo {year} {2016}{\natexlab{b}})}\BibitemShut {NoStop}%
\bibitem [{\citenamefont {Albedah}\ \emph {et~al.}(2018)\citenamefont
  {Albedah}, \citenamefont {Nejadsattari}, \citenamefont {Stadnik},
  \citenamefont {Liu},\ and\ \citenamefont {Cao}}]{Eu1144.Mossbauer}%
  \BibitemOpen
  \bibfield  {author} {\bibinfo {author} {\bibfnamefont {M.~A.}\ \bibnamefont
  {Albedah}}, \bibinfo {author} {\bibfnamefont {F.}~\bibnamefont
  {Nejadsattari}}, \bibinfo {author} {\bibfnamefont {Z.~M.}\ \bibnamefont
  {Stadnik}}, \bibinfo {author} {\bibfnamefont {Y.}~\bibnamefont {Liu}},\ and\
  \bibinfo {author} {\bibfnamefont {G.-H.}\ \bibnamefont {Cao}},\ }\bibinfo
  {title} {M{\"{o}}ssbauer spectroscopy measurements on the 35.5 K
  superconductor
  ${\mathrm{Rb}}_{1\ensuremath{-}\ensuremath{\delta}}{\mathrm{EuFe}}_{4}{\mathrm{As}}_{4}$},\
  \href {https://doi.org/10.1103/PhysRevB.97.144426} {\bibfield  {journal}
  {\bibinfo  {journal} {Phys. Rev. B}\ }\textbf {\bibinfo {volume} {97}},\
  \bibinfo {pages} {144426} (\bibinfo {year} {2018})}\BibitemShut {NoStop}%
\bibitem [{\citenamefont {Iida}\ \emph {et~al.}(2019)\citenamefont {Iida},
  \citenamefont {Nagai}, \citenamefont {Ishida}, \citenamefont {Ishikado},
  \citenamefont {Murai}, \citenamefont {Christianson}, \citenamefont {Yoshida},
  \citenamefont {Inamura}, \citenamefont {Nakamura}, \citenamefont {Nakao},
  \citenamefont {Munakata}, \citenamefont {Kagerbauer}, \citenamefont
  {Eisterer}, \citenamefont {Kawashima}, \citenamefont {Yoshida}, \citenamefont
  {Eisaki},\ and\ \citenamefont {Iyo}}]{Eu1144.neutron}%
  \BibitemOpen
  \bibfield  {author} {\bibinfo {author} {\bibfnamefont {K.}~\bibnamefont
  {Iida}}, \bibinfo {author} {\bibfnamefont {Y.}~\bibnamefont {Nagai}},
  \bibinfo {author} {\bibfnamefont {S.}~\bibnamefont {Ishida}}, \bibinfo
  {author} {\bibfnamefont {M.}~\bibnamefont {Ishikado}}, \bibinfo {author}
  {\bibfnamefont {N.}~\bibnamefont {Murai}}, \bibinfo {author} {\bibfnamefont
  {A.~D.}\ \bibnamefont {Christianson}}, \bibinfo {author} {\bibfnamefont
  {H.}~\bibnamefont {Yoshida}}, \bibinfo {author} {\bibfnamefont
  {Y.}~\bibnamefont {Inamura}}, \bibinfo {author} {\bibfnamefont
  {H.}~\bibnamefont {Nakamura}}, \bibinfo {author} {\bibfnamefont
  {A.}~\bibnamefont {Nakao}}, \bibinfo {author} {\bibfnamefont
  {K.}~\bibnamefont {Munakata}}, \bibinfo {author} {\bibfnamefont
  {D.}~\bibnamefont {Kagerbauer}}, \bibinfo {author} {\bibfnamefont
  {M.}~\bibnamefont {Eisterer}}, \bibinfo {author} {\bibfnamefont
  {K.}~\bibnamefont {Kawashima}}, \bibinfo {author} {\bibfnamefont
  {Y.}~\bibnamefont {Yoshida}}, \bibinfo {author} {\bibfnamefont
  {H.}~\bibnamefont {Eisaki}},\ and\ \bibinfo {author} {\bibfnamefont
  {A.}~\bibnamefont {Iyo}},\ }\bibinfo {title} {Coexisting spin resonance and
  long-range magnetic order of Eu in
  ${\mathrm{EuRbFe}}_{4}{\mathrm{As}}_{4}$},\ \href
  {https://doi.org/10.1103/PhysRevB.100.014506} {\bibfield  {journal} {\bibinfo
   {journal} {Phys. Rev. B}\ }\textbf {\bibinfo {volume} {100}},\ \bibinfo
  {pages} {014506} (\bibinfo {year} {2019})}\BibitemShut {NoStop}%
\bibitem [{\citenamefont {Liu}\ \emph {et~al.}(2017)\citenamefont {Liu},
  \citenamefont {Liu}, \citenamefont {Yu}, \citenamefont {Tao}, \citenamefont
  {Feng},\ and\ \citenamefont {Cao}}]{ly.1144Ni}%
  \BibitemOpen
  \bibfield  {author} {\bibinfo {author} {\bibfnamefont {Y.}~\bibnamefont
  {Liu}}, \bibinfo {author} {\bibfnamefont {Y.-B.}\ \bibnamefont {Liu}},
  \bibinfo {author} {\bibfnamefont {Y.-L.}\ \bibnamefont {Yu}}, \bibinfo
  {author} {\bibfnamefont {Q.}~\bibnamefont {Tao}}, \bibinfo {author}
  {\bibfnamefont {C.-M.}\ \bibnamefont {Feng}},\ and\ \bibinfo {author}
  {\bibfnamefont {G.-H.}\ \bibnamefont {Cao}},\ }\bibinfo {title}
  {$\mathrm{RbEu}{({\mathrm{Fe}}_{1\ensuremath{-}x}{\mathrm{Ni}}_{x})}_{4}{\mathrm{As}}_{4}$:
  From a ferromagnetic superconductor to a superconducting ferromagnet},\ \href
  {https://doi.org/10.1103/PhysRevB.96.224510} {\bibfield  {journal} {\bibinfo
  {journal} {Phys. Rev. B}\ }\textbf {\bibinfo {volume} {96}},\ \bibinfo
  {pages} {224510} (\bibinfo {year} {2017})}\BibitemShut {NoStop}%
\bibitem [{\citenamefont {Liu}\ \emph {et~al.}(2020)\citenamefont {Liu},
  \citenamefont {Liu}, \citenamefont {Cui}, \citenamefont {Ren},\ and\
  \citenamefont {Cao}}]{lyb.1144Co}%
  \BibitemOpen
  \bibfield  {author} {\bibinfo {author} {\bibfnamefont {Y.-B.}\ \bibnamefont
  {Liu}}, \bibinfo {author} {\bibfnamefont {Y.}~\bibnamefont {Liu}}, \bibinfo
  {author} {\bibfnamefont {Y.-W.}\ \bibnamefont {Cui}}, \bibinfo {author}
  {\bibfnamefont {Z.}~\bibnamefont {Ren}},\ and\ \bibinfo {author}
  {\bibfnamefont {G.-H.}\ \bibnamefont {Cao}},\ }\bibinfo {title}
  {Superconductivity and magnetism in
  $\mathrm{RbEu}{({\mathrm{Fe}}_{1\ensuremath{-}x}{\mathrm{Co}}_{x})}_{4}{\mathrm{As}}_{4}$},\
  \href {https://doi.org/10.1088/1361-648x/ab68f4} {\bibfield  {journal}
  {\bibinfo  {journal} {J. Phys.: Condens. Matt.}\ }\textbf {\bibinfo {volume}
  {32}},\ \bibinfo {pages} {175701} (\bibinfo {year} {2020})}\BibitemShut
  {NoStop}%
\bibitem [{\citenamefont {Jiao}\ \emph {et~al.}(2017)\citenamefont {Jiao},
  \citenamefont {Tao}, \citenamefont {Ren}, \citenamefont {Liu},\ and\
  \citenamefont {Cao}}]{jiao2017}%
  \BibitemOpen
  \bibfield  {author} {\bibinfo {author} {\bibfnamefont {W.-H.}\ \bibnamefont
  {Jiao}}, \bibinfo {author} {\bibfnamefont {Q.}~\bibnamefont {Tao}}, \bibinfo
  {author} {\bibfnamefont {Z.}~\bibnamefont {Ren}}, \bibinfo {author}
  {\bibfnamefont {Y.}~\bibnamefont {Liu}},\ and\ \bibinfo {author}
  {\bibfnamefont {G.-H.}\ \bibnamefont {Cao}},\ }\bibinfo {title} {Evidence of
  spontaneous vortex ground state in an iron-based ferromagnetic
  superconductor},\ \href {https://doi.org/10.1038/s41535-017-0057-0}
  {\bibfield  {journal} {\bibinfo  {journal} {npj Quantum Materials}\ }\textbf
  {\bibinfo {volume} {2}},\ \bibinfo {pages} {50} (\bibinfo {year}
  {2017})}\BibitemShut {NoStop}%
\bibitem [{\citenamefont {Vlasko-Vlasov}\ \emph {et~al.}(2019)\citenamefont
  {Vlasko-Vlasov}, \citenamefont {Koshelev}, \citenamefont {Smylie},
  \citenamefont {Bao}, \citenamefont {Chung}, \citenamefont {Kanatzidis},
  \citenamefont {Welp},\ and\ \citenamefont {Kwok}}]{Eu1144.self-flux}%
  \BibitemOpen
  \bibfield  {author} {\bibinfo {author} {\bibfnamefont {V.~K.}\ \bibnamefont
  {Vlasko-Vlasov}}, \bibinfo {author} {\bibfnamefont {A.~E.}\ \bibnamefont
  {Koshelev}}, \bibinfo {author} {\bibfnamefont {M.}~\bibnamefont {Smylie}},
  \bibinfo {author} {\bibfnamefont {J.-K.}\ \bibnamefont {Bao}}, \bibinfo
  {author} {\bibfnamefont {D.~Y.}\ \bibnamefont {Chung}}, \bibinfo {author}
  {\bibfnamefont {M.~G.}\ \bibnamefont {Kanatzidis}}, \bibinfo {author}
  {\bibfnamefont {U.}~\bibnamefont {Welp}},\ and\ \bibinfo {author}
  {\bibfnamefont {W.-K.}\ \bibnamefont {Kwok}},\ }\bibinfo {title}
  {Self-induced magnetic flux structure in the magnetic superconductor
  {${\mathrm{RbEuFe}}_{4}{\mathrm{As}}_{4}$}},\ \href
  {https://doi.org/10.1103/PhysRevB.99.134503} {\bibfield  {journal} {\bibinfo
  {journal} {Phys. Rev. B}\ }\textbf {\bibinfo {volume} {99}},\ \bibinfo
  {pages} {134503} (\bibinfo {year} {2019})}\BibitemShut {NoStop}%
\bibitem [{\citenamefont {Stolyarov}\ \emph {et~al.}(2018)\citenamefont
  {Stolyarov}, \citenamefont {Veshchunov}, \citenamefont {Grebenchuk},
  \citenamefont {Baranov}, \citenamefont {Golovchanskiy}, \citenamefont
  {Shishkin}, \citenamefont {Zhou}, \citenamefont {Shi}, \citenamefont {Xu},
  \citenamefont {Pyon}, \citenamefont {Sun}, \citenamefont {Jiao},
  \citenamefont {Cao}, \citenamefont {Vinnikov}, \citenamefont {Golubov},
  \citenamefont {Tamegai}, \citenamefont {Buzdin},\ and\ \citenamefont
  {Roditchev}}]{Eu122P.DMS}%
  \BibitemOpen
  \bibfield  {author} {\bibinfo {author} {\bibfnamefont {V.~S.}\ \bibnamefont
  {Stolyarov}}, \bibinfo {author} {\bibfnamefont {I.~S.}\ \bibnamefont
  {Veshchunov}}, \bibinfo {author} {\bibfnamefont {S.~Y.}\ \bibnamefont
  {Grebenchuk}}, \bibinfo {author} {\bibfnamefont {D.~S.}\ \bibnamefont
  {Baranov}}, \bibinfo {author} {\bibfnamefont {I.~A.}\ \bibnamefont
  {Golovchanskiy}}, \bibinfo {author} {\bibfnamefont {A.~G.}\ \bibnamefont
  {Shishkin}}, \bibinfo {author} {\bibfnamefont {N.}~\bibnamefont {Zhou}},
  \bibinfo {author} {\bibfnamefont {Z.}~\bibnamefont {Shi}}, \bibinfo {author}
  {\bibfnamefont {X.}~\bibnamefont {Xu}}, \bibinfo {author} {\bibfnamefont
  {S.}~\bibnamefont {Pyon}}, \bibinfo {author} {\bibfnamefont {Y.}~\bibnamefont
  {Sun}}, \bibinfo {author} {\bibfnamefont {W.}~\bibnamefont {Jiao}}, \bibinfo
  {author} {\bibfnamefont {G.-H.}\ \bibnamefont {Cao}}, \bibinfo {author}
  {\bibfnamefont {L.~Y.}\ \bibnamefont {Vinnikov}}, \bibinfo {author}
  {\bibfnamefont {A.~A.}\ \bibnamefont {Golubov}}, \bibinfo {author}
  {\bibfnamefont {T.}~\bibnamefont {Tamegai}}, \bibinfo {author} {\bibfnamefont
  {A.~I.}\ \bibnamefont {Buzdin}},\ and\ \bibinfo {author} {\bibfnamefont
  {D.}~\bibnamefont {Roditchev}},\ }\bibinfo {title} {Domain Meissner state and
  spontaneous vortex-antivortex generation in the ferromagnetic superconductor
  {EuFe$_2$(As$_{0.79}$P$_{0.21}$)$_2$}},\ \href
  {https://doi.org/10.1126/sciadv.aat1061} {\bibfield  {journal} {\bibinfo
  {journal} {Sci. Adv.}\ }\textbf {\bibinfo {volume} {4}},\ \bibinfo {pages}
  {eaat1061} (\bibinfo {year} {2018})}\BibitemShut {NoStop}%
\bibitem [{\citenamefont {Grebenchuk}\ \emph {et~al.}(2020)\citenamefont
  {Grebenchuk}, \citenamefont {Devizorova}, \citenamefont {Golovchanskiy},
  \citenamefont {Shchetinin}, \citenamefont {Cao}, \citenamefont {Buzdin},
  \citenamefont {Roditchev},\ and\ \citenamefont {Stolyarov}}]{Eu122P.SCFM}%
  \BibitemOpen
  \bibfield  {author} {\bibinfo {author} {\bibfnamefont {S.~Y.}\ \bibnamefont
  {Grebenchuk}}, \bibinfo {author} {\bibfnamefont {Z.~A.}\ \bibnamefont
  {Devizorova}}, \bibinfo {author} {\bibfnamefont {I.~A.}\ \bibnamefont
  {Golovchanskiy}}, \bibinfo {author} {\bibfnamefont {I.~V.}\ \bibnamefont
  {Shchetinin}}, \bibinfo {author} {\bibfnamefont {G.-H.}\ \bibnamefont {Cao}},
  \bibinfo {author} {\bibfnamefont {A.~I.}\ \bibnamefont {Buzdin}}, \bibinfo
  {author} {\bibfnamefont {D.}~\bibnamefont {Roditchev}},\ and\ \bibinfo
  {author} {\bibfnamefont {V.~S.}\ \bibnamefont {Stolyarov}},\ }\bibinfo
  {title} {Crossover from ferromagnetic superconductor to superconducting
  ferromagnet in P-doped
  $\mathrm{Eu}{\mathrm{Fe}}_{2}{({\mathrm{As}}_{1\ensuremath{-}x}{\mathrm{P}}_{x})}_{2}$},\
  \href {https://doi.org/10.1103/PhysRevB.102.144501} {\bibfield  {journal}
  {\bibinfo  {journal} {Phys. Rev. B}\ }\textbf {\bibinfo {volume} {102}},\
  \bibinfo {pages} {144501} (\bibinfo {year} {2020})}\BibitemShut {NoStop}%
\bibitem [{\citenamefont {Yu}\ \emph {et~al.}(2017)\citenamefont {Yu},
  \citenamefont {Liu}, \citenamefont {Pan}, \citenamefont {Ruan}, \citenamefont
  {Wang}, \citenamefont {Mu}, \citenamefont {Zhao}, \citenamefont {Chen},\ and\
  \citenamefont {Ren}}]{yu2017}%
  \BibitemOpen
  \bibfield  {author} {\bibinfo {author} {\bibfnamefont {J.}~\bibnamefont
  {Yu}}, \bibinfo {author} {\bibfnamefont {T.}~\bibnamefont {Liu}}, \bibinfo
  {author} {\bibfnamefont {B.-J.}\ \bibnamefont {Pan}}, \bibinfo {author}
  {\bibfnamefont {B.-B.}\ \bibnamefont {Ruan}}, \bibinfo {author}
  {\bibfnamefont {X.-C.}\ \bibnamefont {Wang}}, \bibinfo {author}
  {\bibfnamefont {Q.-G.}\ \bibnamefont {Mu}}, \bibinfo {author} {\bibfnamefont
  {K.}~\bibnamefont {Zhao}}, \bibinfo {author} {\bibfnamefont {G.-F.}\
  \bibnamefont {Chen}},\ and\ \bibinfo {author} {\bibfnamefont {Z.-A.}\
  \bibnamefont {Ren}},\ }\bibinfo {title} {Discovery of a novel 112-type
  iron-pnictide and La-doping induced superconductivity in
  Eu$_{1-x}$La$_{x}$FeAs$_{2}$ ($x$ = 0-0.15)},\ \href
  {https://doi.org/10.1016/j.scib.2016.12.015} {\bibfield  {journal} {\bibinfo
  {journal} {Sci. Bull.}\ }\textbf {\bibinfo {volume} {62}},\ \bibinfo {pages}
  {218} (\bibinfo {year} {2017})}\BibitemShut {NoStop}%
\bibitem [{\citenamefont {Katayama}\ \emph {et~al.}(2013)\citenamefont
  {Katayama}, \citenamefont {Kudo}, \citenamefont {Onari}, \citenamefont
  {Mizukami}, \citenamefont {Sugawara}, \citenamefont {Sugiyama}, \citenamefont
  {Kitahama}, \citenamefont {Iba}, \citenamefont {Fujimura}, \citenamefont
  {Nishimoto}, \citenamefont {Nohara},\ and\ \citenamefont
  {Sawa}}]{katayama2013}%
  \BibitemOpen
  \bibfield  {author} {\bibinfo {author} {\bibfnamefont {N.}~\bibnamefont
  {Katayama}}, \bibinfo {author} {\bibfnamefont {K.}~\bibnamefont {Kudo}},
  \bibinfo {author} {\bibfnamefont {S.}~\bibnamefont {Onari}}, \bibinfo
  {author} {\bibfnamefont {T.}~\bibnamefont {Mizukami}}, \bibinfo {author}
  {\bibfnamefont {K.}~\bibnamefont {Sugawara}}, \bibinfo {author}
  {\bibfnamefont {Y.}~\bibnamefont {Sugiyama}}, \bibinfo {author}
  {\bibfnamefont {Y.}~\bibnamefont {Kitahama}}, \bibinfo {author}
  {\bibfnamefont {K.}~\bibnamefont {Iba}}, \bibinfo {author} {\bibfnamefont
  {K.}~\bibnamefont {Fujimura}}, \bibinfo {author} {\bibfnamefont
  {N.}~\bibnamefont {Nishimoto}}, \bibinfo {author} {\bibfnamefont
  {M.}~\bibnamefont {Nohara}},\ and\ \bibinfo {author} {\bibfnamefont
  {H.}~\bibnamefont {Sawa}},\ }\bibinfo {title} {Superconductivity in
  Ca$_{1-x}$La$_{x}$FeAs$_{2}$: A Novel 112-Type Iron Pnictide with Arsenic
  Zigzag Bonds},\ \href {https://doi.org/10.7566/JPSJ.82.123702} {\bibfield
  {journal} {\bibinfo  {journal} {J. Phys. Soc. Jpn.}\ }\textbf {\bibinfo
  {volume} {82}},\ \bibinfo {pages} {123702} (\bibinfo {year}
  {2013})}\BibitemShut {NoStop}%
\bibitem [{\citenamefont {Yakita}\ \emph {et~al.}(2014)\citenamefont {Yakita},
  \citenamefont {Ogino}, \citenamefont {Okada}, \citenamefont {Yamamoto},
  \citenamefont {Kishio}, \citenamefont {Tohei}, \citenamefont {Ikuhara},
  \citenamefont {Gotoh}, \citenamefont {Fujihisa}, \citenamefont {Kataoka},
  \citenamefont {Eisaki},\ and\ \citenamefont {Shimoyama}}]{yakita2014}%
  \BibitemOpen
  \bibfield  {author} {\bibinfo {author} {\bibfnamefont {H.}~\bibnamefont
  {Yakita}}, \bibinfo {author} {\bibfnamefont {H.}~\bibnamefont {Ogino}},
  \bibinfo {author} {\bibfnamefont {T.}~\bibnamefont {Okada}}, \bibinfo
  {author} {\bibfnamefont {A.}~\bibnamefont {Yamamoto}}, \bibinfo {author}
  {\bibfnamefont {K.}~\bibnamefont {Kishio}}, \bibinfo {author} {\bibfnamefont
  {T.}~\bibnamefont {Tohei}}, \bibinfo {author} {\bibfnamefont
  {Y.}~\bibnamefont {Ikuhara}}, \bibinfo {author} {\bibfnamefont
  {Y.}~\bibnamefont {Gotoh}}, \bibinfo {author} {\bibfnamefont
  {H.}~\bibnamefont {Fujihisa}}, \bibinfo {author} {\bibfnamefont
  {K.}~\bibnamefont {Kataoka}}, \bibinfo {author} {\bibfnamefont
  {H.}~\bibnamefont {Eisaki}},\ and\ \bibinfo {author} {\bibfnamefont {J.-i.}\
  \bibnamefont {Shimoyama}},\ }\bibinfo {title} {A New Layered Iron Arsenide
  Superconductor: (Ca,Pr)FeAs$_{2}$},\ \href
  {https://doi.org/10.1021/ja410845b} {\bibfield  {journal} {\bibinfo
  {journal} {J. Am. Chem. Soc.}\ }\textbf {\bibinfo {volume} {136}},\ \bibinfo
  {pages} {846} (\bibinfo {year} {2014})}\BibitemShut {NoStop}%
\bibitem [{\citenamefont {Sala}\ \emph {et~al.}(2014)\citenamefont {Sala},
  \citenamefont {Yakita}, \citenamefont {Ogino}, \citenamefont {Okada},
  \citenamefont {Yamamoto}, \citenamefont {Kishio}, \citenamefont {Ishida},
  \citenamefont {Iyo}, \citenamefont {Eisaki}, \citenamefont {Fujioka},
  \citenamefont {Takano}, \citenamefont {Putti},\ and\ \citenamefont {ichi
  Shimoyama}}]{Sala2014}%
  \BibitemOpen
  \bibfield  {author} {\bibinfo {author} {\bibfnamefont {A.}~\bibnamefont
  {Sala}}, \bibinfo {author} {\bibfnamefont {H.}~\bibnamefont {Yakita}},
  \bibinfo {author} {\bibfnamefont {H.}~\bibnamefont {Ogino}}, \bibinfo
  {author} {\bibfnamefont {T.}~\bibnamefont {Okada}}, \bibinfo {author}
  {\bibfnamefont {A.}~\bibnamefont {Yamamoto}}, \bibinfo {author}
  {\bibfnamefont {K.}~\bibnamefont {Kishio}}, \bibinfo {author} {\bibfnamefont
  {S.}~\bibnamefont {Ishida}}, \bibinfo {author} {\bibfnamefont
  {A.}~\bibnamefont {Iyo}}, \bibinfo {author} {\bibfnamefont {H.}~\bibnamefont
  {Eisaki}}, \bibinfo {author} {\bibfnamefont {M.}~\bibnamefont {Fujioka}},
  \bibinfo {author} {\bibfnamefont {Y.}~\bibnamefont {Takano}}, \bibinfo
  {author} {\bibfnamefont {M.}~\bibnamefont {Putti}},\ and\ \bibinfo {author}
  {\bibfnamefont {J.}~\bibnamefont {ichi Shimoyama}},\ }\bibinfo {title}
  {Synthesis and physical properties of {Ca$_{1-x}$RE$_x$FeAs$_2$ with RE =
  La{\textendash}Gd}},\ \href {https://doi.org/10.7567/apex.7.073102}
  {\bibfield  {journal} {\bibinfo  {journal} {Appl. Phys. Express}\ }\textbf
  {\bibinfo {volume} {7}},\ \bibinfo {pages} {073102} (\bibinfo {year}
  {2014})}\BibitemShut {NoStop}%
\bibitem [{\citenamefont {Ray}\ and\ \citenamefont {Alff}(2017)}]{Ray2017}%
  \BibitemOpen
  \bibfield  {author} {\bibinfo {author} {\bibfnamefont {S.~J.}\ \bibnamefont
  {Ray}}\ and\ \bibinfo {author} {\bibfnamefont {L.}~\bibnamefont {Alff}},\
  }\bibinfo {title} {Superconductivity and Dirac fermions in 112-phase
  pnictides},\ \href {https://doi.org/https://doi.org/10.1002/pssb.201600163}
  {\bibfield  {journal} {\bibinfo  {journal} {Physica Status Solidi (b)}\
  }\textbf {\bibinfo {volume} {254}},\ \bibinfo {pages} {1600163} (\bibinfo
  {year} {2017})}\BibitemShut {NoStop}%
\bibitem [{\citenamefont {Nohara}\ and\ \citenamefont
  {Kudo}(2017)}]{Nohara2017}%
  \BibitemOpen
  \bibfield  {author} {\bibinfo {author} {\bibfnamefont {M.}~\bibnamefont
  {Nohara}}\ and\ \bibinfo {author} {\bibfnamefont {K.}~\bibnamefont {Kudo}},\
  }\bibinfo {title} {Arsenic chemistry of iron-based superconductors and
  strategy for novel superconducting materials},\ \href
  {https://doi.org/10.1080/23746149.2017.1317024} {\bibfield  {journal}
  {\bibinfo  {journal} {Adv. Phys. X}\ }\textbf {\bibinfo {volume} {2}},\
  \bibinfo {pages} {450} (\bibinfo {year} {2017})}\BibitemShut {NoStop}%
\bibitem [{\citenamefont {Li}\ \emph {et~al.}(2015)\citenamefont {Li},
  \citenamefont {Liu}, \citenamefont {Zhou}, \citenamefont {Yang},
  \citenamefont {Shen}, \citenamefont {Li}, \citenamefont {Jiang},
  \citenamefont {Niu}, \citenamefont {Xie}, \citenamefont {Sun}, \citenamefont
  {Fan}, \citenamefont {Yao}, \citenamefont {Liu}, \citenamefont {Shi},\ and\
  \citenamefont {Xie}}]{li2015.112}%
  \BibitemOpen
  \bibfield  {author} {\bibinfo {author} {\bibfnamefont {M.~Y.}\ \bibnamefont
  {Li}}, \bibinfo {author} {\bibfnamefont {Z.~T.}\ \bibnamefont {Liu}},
  \bibinfo {author} {\bibfnamefont {W.}~\bibnamefont {Zhou}}, \bibinfo {author}
  {\bibfnamefont {H.~F.}\ \bibnamefont {Yang}}, \bibinfo {author}
  {\bibfnamefont {D.~W.}\ \bibnamefont {Shen}}, \bibinfo {author}
  {\bibfnamefont {W.}~\bibnamefont {Li}}, \bibinfo {author} {\bibfnamefont
  {J.}~\bibnamefont {Jiang}}, \bibinfo {author} {\bibfnamefont {X.~H.}\
  \bibnamefont {Niu}}, \bibinfo {author} {\bibfnamefont {B.~P.}\ \bibnamefont
  {Xie}}, \bibinfo {author} {\bibfnamefont {Y.}~\bibnamefont {Sun}}, \bibinfo
  {author} {\bibfnamefont {C.~C.}\ \bibnamefont {Fan}}, \bibinfo {author}
  {\bibfnamefont {Q.}~\bibnamefont {Yao}}, \bibinfo {author} {\bibfnamefont
  {J.~S.}\ \bibnamefont {Liu}}, \bibinfo {author} {\bibfnamefont {Z.~X.}\
  \bibnamefont {Shi}},\ and\ \bibinfo {author} {\bibfnamefont {X.~M.}\
  \bibnamefont {Xie}},\ }\bibinfo {title} {Significant contribution of As $4p$
  orbitals to the low-lying electronic structure of the 112-type iron-based
  superconductor $\mathrm{Ca}{}_{0.9}\mathrm{La}{}_{0.1}\mathrm{FeAs}{}_{2}$},\
  \href {https://doi.org/10.1103/PhysRevB.91.045112} {\bibfield  {journal}
  {\bibinfo  {journal} {Phys. Rev. B}\ }\textbf {\bibinfo {volume} {91}},\
  \bibinfo {pages} {045112} (\bibinfo {year} {2015})}\BibitemShut {NoStop}%
\bibitem [{\citenamefont {Jiang}\ \emph {et~al.}(2016)\citenamefont {Jiang},
  \citenamefont {Liu}, \citenamefont {Sch\"utt}, \citenamefont {Hallas},
  \citenamefont {Shen}, \citenamefont {Tian}, \citenamefont {Emmanouilidou},
  \citenamefont {Shi}, \citenamefont {Luke}, \citenamefont {Uemura},
  \citenamefont {Fernandes},\ and\ \citenamefont {Ni}}]{jiang2016.112}%
  \BibitemOpen
  \bibfield  {author} {\bibinfo {author} {\bibfnamefont {S.}~\bibnamefont
  {Jiang}}, \bibinfo {author} {\bibfnamefont {L.}~\bibnamefont {Liu}}, \bibinfo
  {author} {\bibfnamefont {M.}~\bibnamefont {Sch\"utt}}, \bibinfo {author}
  {\bibfnamefont {A.~M.}\ \bibnamefont {Hallas}}, \bibinfo {author}
  {\bibfnamefont {B.}~\bibnamefont {Shen}}, \bibinfo {author} {\bibfnamefont
  {W.}~\bibnamefont {Tian}}, \bibinfo {author} {\bibfnamefont {E.}~\bibnamefont
  {Emmanouilidou}}, \bibinfo {author} {\bibfnamefont {A.}~\bibnamefont {Shi}},
  \bibinfo {author} {\bibfnamefont {G.~M.}\ \bibnamefont {Luke}}, \bibinfo
  {author} {\bibfnamefont {Y.~J.}\ \bibnamefont {Uemura}}, \bibinfo {author}
  {\bibfnamefont {R.~M.}\ \bibnamefont {Fernandes}},\ and\ \bibinfo {author}
  {\bibfnamefont {N.}~\bibnamefont {Ni}},\ }\bibinfo {title} {Effect of
  interlayer coupling on the coexistence of antiferromagnetism and
  superconductivity in Fe pnictide superconductors: A study of
  ${\mathrm{Ca}}_{0.74(1)}{\mathrm{La}}_{0.26(1)}({\mathrm{Fe}}_{1\ensuremath{-}x}{\mathrm{Co}}_{x}){\mathrm{As}}_{2}$
  single crystals},\ \href {https://doi.org/10.1103/PhysRevB.93.174513}
  {\bibfield  {journal} {\bibinfo  {journal} {Phys. Rev. B}\ }\textbf {\bibinfo
  {volume} {93}},\ \bibinfo {pages} {174513} (\bibinfo {year}
  {2016})}\BibitemShut {NoStop}%
\bibitem [{\citenamefont {Wu}\ \emph {et~al.}(2014)\citenamefont {Wu},
  \citenamefont {Le}, \citenamefont {Liang}, \citenamefont {Qin}, \citenamefont
  {Fan},\ and\ \citenamefont {Hu}}]{wuxx.PRB2014}%
  \BibitemOpen
  \bibfield  {author} {\bibinfo {author} {\bibfnamefont {X.}~\bibnamefont
  {Wu}}, \bibinfo {author} {\bibfnamefont {C.}~\bibnamefont {Le}}, \bibinfo
  {author} {\bibfnamefont {Y.}~\bibnamefont {Liang}}, \bibinfo {author}
  {\bibfnamefont {S.}~\bibnamefont {Qin}}, \bibinfo {author} {\bibfnamefont
  {H.}~\bibnamefont {Fan}},\ and\ \bibinfo {author} {\bibfnamefont
  {J.}~\bibnamefont {Hu}},\ }\bibinfo {title} {Effect of As-chain layers in
  ${\text{CaFeAs}}_{2}$},\ \href {https://doi.org/10.1103/PhysRevB.89.205102}
  {\bibfield  {journal} {\bibinfo  {journal} {Phys. Rev. B}\ }\textbf {\bibinfo
  {volume} {89}},\ \bibinfo {pages} {205102} (\bibinfo {year}
  {2014})}\BibitemShut {NoStop}%
\bibitem [{\citenamefont {Wu}\ \emph {et~al.}(2015)\citenamefont {Wu},
  \citenamefont {Qin}, \citenamefont {Liang}, \citenamefont {Le}, \citenamefont
  {Fan},\ and\ \citenamefont {Hu}}]{wuxx.prb2015}%
  \BibitemOpen
  \bibfield  {author} {\bibinfo {author} {\bibfnamefont {X.}~\bibnamefont
  {Wu}}, \bibinfo {author} {\bibfnamefont {S.}~\bibnamefont {Qin}}, \bibinfo
  {author} {\bibfnamefont {Y.}~\bibnamefont {Liang}}, \bibinfo {author}
  {\bibfnamefont {C.}~\bibnamefont {Le}}, \bibinfo {author} {\bibfnamefont
  {H.}~\bibnamefont {Fan}},\ and\ \bibinfo {author} {\bibfnamefont
  {J.}~\bibnamefont {Hu}},\ }\bibinfo {title} {${\mathrm{CaFeAs}}_{2}$: A
  staggered intercalation of quantum spin Hall and high-temperature
  superconductivity},\ \href {https://doi.org/10.1103/PhysRevB.91.081111}
  {\bibfield  {journal} {\bibinfo  {journal} {Phys. Rev. B}\ }\textbf {\bibinfo
  {volume} {91}},\ \bibinfo {pages} {081111} (\bibinfo {year}
  {2015})}\BibitemShut {NoStop}%
\bibitem [{\citenamefont {Liu}\ \emph {et~al.}(2016{\natexlab{c}})\citenamefont
  {Liu}, \citenamefont {Xing}, \citenamefont {Li}, \citenamefont {Zhou},
  \citenamefont {Sun}, \citenamefont {Fan}, \citenamefont {Yang}, \citenamefont
  {Liu}, \citenamefont {Yao}, \citenamefont {Li}, \citenamefont {Shi},
  \citenamefont {Shen},\ and\ \citenamefont {Wang}}]{Dirac.APL2016}%
  \BibitemOpen
  \bibfield  {author} {\bibinfo {author} {\bibfnamefont {Z.~T.}\ \bibnamefont
  {Liu}}, \bibinfo {author} {\bibfnamefont {X.~Z.}\ \bibnamefont {Xing}},
  \bibinfo {author} {\bibfnamefont {M.~Y.}\ \bibnamefont {Li}}, \bibinfo
  {author} {\bibfnamefont {W.}~\bibnamefont {Zhou}}, \bibinfo {author}
  {\bibfnamefont {Y.}~\bibnamefont {Sun}}, \bibinfo {author} {\bibfnamefont
  {C.~C.}\ \bibnamefont {Fan}}, \bibinfo {author} {\bibfnamefont {H.~F.}\
  \bibnamefont {Yang}}, \bibinfo {author} {\bibfnamefont {J.~S.}\ \bibnamefont
  {Liu}}, \bibinfo {author} {\bibfnamefont {Q.}~\bibnamefont {Yao}}, \bibinfo
  {author} {\bibfnamefont {W.}~\bibnamefont {Li}}, \bibinfo {author}
  {\bibfnamefont {Z.~X.}\ \bibnamefont {Shi}}, \bibinfo {author} {\bibfnamefont
  {D.~W.}\ \bibnamefont {Shen}},\ and\ \bibinfo {author} {\bibfnamefont
  {Z.}~\bibnamefont {Wang}},\ }\bibinfo {title} {Observation of the anisotropic
  Dirac cone in the band dispersion of 112-structured iron-based superconductor
  {Ca$_{0.9}$La$_{0.1}$FeAs$_{2}$}},\ \href {https://doi.org/10.1063/1.4960164}
  {\bibfield  {journal} {\bibinfo  {journal} {Appl. Phys. Lett.}\ }\textbf
  {\bibinfo {volume} {109}},\ \bibinfo {pages} {042602} (\bibinfo {year}
  {2016}{\natexlab{c}})}\BibitemShut {NoStop}%
\bibitem [{\citenamefont {Wu}\ \emph {et~al.}(2020)\citenamefont {Wu},
  \citenamefont {Benalcazar}, \citenamefont {Li}, \citenamefont {Thomale},
  \citenamefont {Liu},\ and\ \citenamefont {Hu}}]{wuxx.PRX2020}%
  \BibitemOpen
  \bibfield  {author} {\bibinfo {author} {\bibfnamefont {X.}~\bibnamefont
  {Wu}}, \bibinfo {author} {\bibfnamefont {W.~A.}\ \bibnamefont {Benalcazar}},
  \bibinfo {author} {\bibfnamefont {Y.}~\bibnamefont {Li}}, \bibinfo {author}
  {\bibfnamefont {R.}~\bibnamefont {Thomale}}, \bibinfo {author} {\bibfnamefont
  {C.-X.}\ \bibnamefont {Liu}},\ and\ \bibinfo {author} {\bibfnamefont
  {J.}~\bibnamefont {Hu}},\ }\bibinfo {title} {Boundary-Obstructed Topological
  High-${\mathit{T}}_{c}$ Superconductivity in Iron Pnictides},\ \href
  {https://doi.org/10.1103/PhysRevX.10.041014} {\bibfield  {journal} {\bibinfo
  {journal} {Phys. Rev. X}\ }\textbf {\bibinfo {volume} {10}},\ \bibinfo
  {pages} {041014} (\bibinfo {year} {2020})}\BibitemShut {NoStop}%
\bibitem [{\citenamefont {Albedah}\ \emph {et~al.}(2020)\citenamefont
  {Albedah}, \citenamefont {Stadnik}, \citenamefont {Fedoryk}, \citenamefont
  {Liu},\ and\ \citenamefont {Cao}}]{albedah2020}%
  \BibitemOpen
  \bibfield  {author} {\bibinfo {author} {\bibfnamefont {M.~A.}\ \bibnamefont
  {Albedah}}, \bibinfo {author} {\bibfnamefont {Z.~M.}\ \bibnamefont
  {Stadnik}}, \bibinfo {author} {\bibfnamefont {O.}~\bibnamefont {Fedoryk}},
  \bibinfo {author} {\bibfnamefont {Y.-B.}\ \bibnamefont {Liu}},\ and\ \bibinfo
  {author} {\bibfnamefont {G.-H.}\ \bibnamefont {Cao}},\ }\bibinfo {title}
  {Magnetic properties of EuFeAs$_{2}$ and the 14 K superconductor
  EuFe$_{0.97}$Ni$_{0.03}$As$_{2}$},\ \href
  {https://doi.org/https://doi.org/10.1016/j.jmmm.2020.166603} {\bibfield
  {journal} {\bibinfo  {journal} {J. Magn. Magn. Mater.}\ }\textbf {\bibinfo
  {volume} {503}},\ \bibinfo {pages} {166603} (\bibinfo {year}
  {2020})}\BibitemShut {NoStop}%
\bibitem [{\citenamefont {Liu}\ \emph {et~al.}(2018)\citenamefont {Liu},
  \citenamefont {Liu}, \citenamefont {Jiao}, \citenamefont {Ren},\ and\
  \citenamefont {Cao}}]{liuyb2018}%
  \BibitemOpen
  \bibfield  {author} {\bibinfo {author} {\bibfnamefont {Y.}~\bibnamefont
  {Liu}}, \bibinfo {author} {\bibfnamefont {Y.}~\bibnamefont {Liu}}, \bibinfo
  {author} {\bibfnamefont {W.}~\bibnamefont {Jiao}}, \bibinfo {author}
  {\bibfnamefont {Z.}~\bibnamefont {Ren}},\ and\ \bibinfo {author}
  {\bibfnamefont {G.}~\bibnamefont {Cao}},\ }\bibinfo {title} {Magnetism and
  superconductivity in Eu(Fe$_{1-x}$Ni$_{x}$)As$_{2}$ ($x$ = 0, 0.04)},\ \href
  {https://doi.org/10.1007/s11433-018-9284-3} {\bibfield  {journal} {\bibinfo
  {journal} {Sci. China Phys. Mech. Astron.}\ }\textbf {\bibinfo {volume}
  {61}},\ \bibinfo {pages} {127405} (\bibinfo {year} {2018})}\BibitemShut
  {NoStop}%
\bibitem [{\citenamefont {Yakita}\ \emph {et~al.}(2015)\citenamefont {Yakita},
  \citenamefont {Ogino}, \citenamefont {Sala}, \citenamefont {Okada},
  \citenamefont {Yamamoto}, \citenamefont {Kishio}, \citenamefont {Iyo},
  \citenamefont {Eisaki},\ and\ \citenamefont {ichi Shimoyama}}]{Yakita2015}%
  \BibitemOpen
  \bibfield  {author} {\bibinfo {author} {\bibfnamefont {H.}~\bibnamefont
  {Yakita}}, \bibinfo {author} {\bibfnamefont {H.}~\bibnamefont {Ogino}},
  \bibinfo {author} {\bibfnamefont {A.}~\bibnamefont {Sala}}, \bibinfo {author}
  {\bibfnamefont {T.}~\bibnamefont {Okada}}, \bibinfo {author} {\bibfnamefont
  {A.}~\bibnamefont {Yamamoto}}, \bibinfo {author} {\bibfnamefont
  {K.}~\bibnamefont {Kishio}}, \bibinfo {author} {\bibfnamefont
  {A.}~\bibnamefont {Iyo}}, \bibinfo {author} {\bibfnamefont {H.}~\bibnamefont
  {Eisaki}},\ and\ \bibinfo {author} {\bibfnamefont {J.}~\bibnamefont {ichi
  Shimoyama}},\ }\bibinfo {title} {Co and Mn doping effect in polycrystalline
  {(Ca,La) and (Ca,Pr)FeAs$_2$} superconductors},\ \href
  {https://doi.org/10.1088/0953-2048/28/6/065001} {\bibfield  {journal}
  {\bibinfo  {journal} {Supercond. Sci. Technol.}\ }\textbf {\bibinfo {volume}
  {28}},\ \bibinfo {pages} {065001} (\bibinfo {year} {2015})}\BibitemShut
  {NoStop}%
\bibitem [{\citenamefont {Kang}\ \emph {et~al.}(2017)\citenamefont {Kang},
  \citenamefont {Birol},\ and\ \citenamefont {Kotliar}}]{kotliar.prb2017}%
  \BibitemOpen
  \bibfield  {author} {\bibinfo {author} {\bibfnamefont {C.-J.}\ \bibnamefont
  {Kang}}, \bibinfo {author} {\bibfnamefont {T.}~\bibnamefont {Birol}},\ and\
  \bibinfo {author} {\bibfnamefont {G.}~\bibnamefont {Kotliar}},\ }\bibinfo
  {title} {Phase stability and large in-plane resistivity anisotropy in the
  112-type iron-based superconductor
  ${\mathrm{Ca}}_{1\ensuremath{-}x}{\mathrm{La}}_{x}{\mathrm{FeAs}}_{2}$},\
  \href {https://doi.org/10.1103/PhysRevB.95.014511} {\bibfield  {journal}
  {\bibinfo  {journal} {Phys. Rev. B}\ }\textbf {\bibinfo {volume} {95}},\
  \bibinfo {pages} {014511} (\bibinfo {year} {2017})}\BibitemShut {NoStop}%
\bibitem [{liu()}]{liu.SM}%
  \BibitemOpen
  \href@noop {} {\ }\bibinfo {note} {{Supplemental Material which contains two
  parts: I. XRD Rietveld refinements of Eu(Fe$_{1-x}$Ni$_{x}$)As$_{2}$; II. AC
  magnetic susceptibility measurement.}}\BibitemShut {Stop}%
\bibitem [{\citenamefont {Takikawa}\ \emph {et~al.}(2010)\citenamefont
  {Takikawa}, \citenamefont {Ebisu},\ and\ \citenamefont
  {Nagata}}]{takikawa2010.Eu2O3}%
  \BibitemOpen
  \bibfield  {author} {\bibinfo {author} {\bibfnamefont {Y.}~\bibnamefont
  {Takikawa}}, \bibinfo {author} {\bibfnamefont {S.}~\bibnamefont {Ebisu}},\
  and\ \bibinfo {author} {\bibfnamefont {S.}~\bibnamefont {Nagata}},\ }\bibinfo
  {title} {{Van Vleck paramagnetism of the trivalent Eu ions}},\ \href
  {https://doi.org/https://doi.org/10.1016/j.jpcs.2010.08.006} {\bibfield
  {journal} {\bibinfo  {journal} {J Phys Chem Solids}\ }\textbf {\bibinfo
  {volume} {71}},\ \bibinfo {pages} {1592} (\bibinfo {year}
  {2010})}\BibitemShut {NoStop}%
\bibitem [{\citenamefont {Yuzuri}\ \emph {et~al.}(1980)\citenamefont {Yuzuri},
  \citenamefont {Tahara},\ and\ \citenamefont {Nakamura}}]{yuzuri1980.FeAs2}%
  \BibitemOpen
  \bibfield  {author} {\bibinfo {author} {\bibfnamefont {M.}~\bibnamefont
  {Yuzuri}}, \bibinfo {author} {\bibfnamefont {R.}~\bibnamefont {Tahara}},\
  and\ \bibinfo {author} {\bibfnamefont {Y.}~\bibnamefont {Nakamura}},\
  }\bibinfo {title} {{M\"{o}ssbauer Study of Iron-Arsenic Compounds}},\ \href
  {https://doi.org/10.1143/JPSJ.48.1937} {\bibfield  {journal} {\bibinfo
  {journal} {J. Phys. Soc. Japan}\ }\textbf {\bibinfo {volume} {48}},\ \bibinfo
  {pages} {1937} (\bibinfo {year} {1980})}\BibitemShut {NoStop}%
\bibitem [{\citenamefont {Cao}\ \emph {et~al.}(2009)\citenamefont {Cao},
  \citenamefont {Jiang}, \citenamefont {Lin}, \citenamefont {Wang},
  \citenamefont {Li}, \citenamefont {Ren}, \citenamefont {Tao}, \citenamefont
  {Feng}, \citenamefont {Dai}, \citenamefont {Xu},\ and\ \citenamefont
  {Zhang}}]{cao2009}%
  \BibitemOpen
  \bibfield  {author} {\bibinfo {author} {\bibfnamefont {G.}~\bibnamefont
  {Cao}}, \bibinfo {author} {\bibfnamefont {S.}~\bibnamefont {Jiang}}, \bibinfo
  {author} {\bibfnamefont {X.}~\bibnamefont {Lin}}, \bibinfo {author}
  {\bibfnamefont {C.}~\bibnamefont {Wang}}, \bibinfo {author} {\bibfnamefont
  {Y.}~\bibnamefont {Li}}, \bibinfo {author} {\bibfnamefont {Z.}~\bibnamefont
  {Ren}}, \bibinfo {author} {\bibfnamefont {Q.}~\bibnamefont {Tao}}, \bibinfo
  {author} {\bibfnamefont {C.}~\bibnamefont {Feng}}, \bibinfo {author}
  {\bibfnamefont {J.}~\bibnamefont {Dai}}, \bibinfo {author} {\bibfnamefont
  {Z.}~\bibnamefont {Xu}},\ and\ \bibinfo {author} {\bibfnamefont {F.-C.}\
  \bibnamefont {Zhang}},\ }\bibinfo {title} {Narrow superconducting window in
  ${\text{LaFe}}_{1\ensuremath{-}x}{\text{Ni}}_{x}\text{AsO}$},\ \href
  {https://doi.org/10.1103/PhysRevB.79.174505} {\bibfield  {journal} {\bibinfo
  {journal} {Phys. Rev. B}\ }\textbf {\bibinfo {volume} {79}},\ \bibinfo
  {pages} {174505} (\bibinfo {year} {2009})}\BibitemShut {NoStop}%
\bibitem [{\citenamefont {Ren}\ \emph {et~al.}(2009{\natexlab{b}})\citenamefont
  {Ren}, \citenamefont {Lin}, \citenamefont {Tao}, \citenamefont {Jiang},
  \citenamefont {Zhu}, \citenamefont {Wang}, \citenamefont {Cao},\ and\
  \citenamefont {Xu}}]{ren2009-1}%
  \BibitemOpen
  \bibfield  {author} {\bibinfo {author} {\bibfnamefont {Z.}~\bibnamefont
  {Ren}}, \bibinfo {author} {\bibfnamefont {X.}~\bibnamefont {Lin}}, \bibinfo
  {author} {\bibfnamefont {Q.}~\bibnamefont {Tao}}, \bibinfo {author}
  {\bibfnamefont {S.}~\bibnamefont {Jiang}}, \bibinfo {author} {\bibfnamefont
  {Z.}~\bibnamefont {Zhu}}, \bibinfo {author} {\bibfnamefont {C.}~\bibnamefont
  {Wang}}, \bibinfo {author} {\bibfnamefont {G.}~\bibnamefont {Cao}},\ and\
  \bibinfo {author} {\bibfnamefont {Z.}~\bibnamefont {Xu}},\ }\bibinfo {title}
  {Suppression of spin-density-wave transition and emergence of ferromagnetic
  ordering of ${\text{Eu}}^{2+}$ moments in
  ${\text{EuFe}}_{2\ensuremath{-}x}{\text{Ni}}_{x}{\text{As}}_{2}$},\ \href
  {https://doi.org/10.1103/PhysRevB.79.094426} {\bibfield  {journal} {\bibinfo
  {journal} {Phys. Rev. B}\ }\textbf {\bibinfo {volume} {79}},\ \bibinfo
  {pages} {094426} (\bibinfo {year} {2009}{\natexlab{b}})}\BibitemShut
  {NoStop}%
\bibitem [{\citenamefont {Jiang}\ \emph {et~al.}(2009)\citenamefont {Jiang},
  \citenamefont {Luo}, \citenamefont {Ren}, \citenamefont {Zhu}, \citenamefont
  {Wang}, \citenamefont {Xu}, \citenamefont {Tao}, \citenamefont {Cao},\ and\
  \citenamefont {Xu}}]{jiang2009}%
  \BibitemOpen
  \bibfield  {author} {\bibinfo {author} {\bibfnamefont {S.}~\bibnamefont
  {Jiang}}, \bibinfo {author} {\bibfnamefont {Y.}~\bibnamefont {Luo}}, \bibinfo
  {author} {\bibfnamefont {Z.}~\bibnamefont {Ren}}, \bibinfo {author}
  {\bibfnamefont {Z.}~\bibnamefont {Zhu}}, \bibinfo {author} {\bibfnamefont
  {C.}~\bibnamefont {Wang}}, \bibinfo {author} {\bibfnamefont {X.}~\bibnamefont
  {Xu}}, \bibinfo {author} {\bibfnamefont {Q.}~\bibnamefont {Tao}}, \bibinfo
  {author} {\bibfnamefont {G.}~\bibnamefont {Cao}},\ and\ \bibinfo {author}
  {\bibfnamefont {Z.}~\bibnamefont {Xu}},\ }\bibinfo {title} {Metamagnetic
  transition in {${\mathrm{EuFe}}_{2}{\mathrm{As}}_{2}$} single crystals},\
  \href {http://stacks.iop.org/1367-2630/11/i=2/a=025007} {\bibfield  {journal}
  {\bibinfo  {journal} {New J. Phys.}\ }\textbf {\bibinfo {volume} {11}},\
  \bibinfo {pages} {025007} (\bibinfo {year} {2009})}\BibitemShut {NoStop}%
\bibitem [{\citenamefont {Sangeetha}\ \emph {et~al.}(2019)\citenamefont
  {Sangeetha}, \citenamefont {Smetana}, \citenamefont {Mudring},\ and\
  \citenamefont {Johnston}}]{Eu122Ni.PRB2019}%
  \BibitemOpen
  \bibfield  {author} {\bibinfo {author} {\bibfnamefont {N.~S.}\ \bibnamefont
  {Sangeetha}}, \bibinfo {author} {\bibfnamefont {V.}~\bibnamefont {Smetana}},
  \bibinfo {author} {\bibfnamefont {A.-V.}\ \bibnamefont {Mudring}},\ and\
  \bibinfo {author} {\bibfnamefont {D.~C.}\ \bibnamefont {Johnston}},\
  }\bibinfo {title} {Helical antiferromagnetic ordering in
  ${\mathrm{EuNi}}_{1.95}{\mathrm{As}}_{2}$ single crystals},\ \href
  {https://doi.org/10.1103/PhysRevB.100.094438} {\bibfield  {journal} {\bibinfo
   {journal} {Phys. Rev. B}\ }\textbf {\bibinfo {volume} {100}},\ \bibinfo
  {pages} {094438} (\bibinfo {year} {2019})}\BibitemShut {NoStop}%
\bibitem [{\citenamefont {Jin}\ \emph {et~al.}(2019)\citenamefont {Jin},
  \citenamefont {Qureshi}, \citenamefont {Bukowski}, \citenamefont {Xiao},
  \citenamefont {Nandi}, \citenamefont {Babij}, \citenamefont {Fu},
  \citenamefont {Su},\ and\ \citenamefont {Br\"uckel}}]{jin2019.Eu122Ni}%
  \BibitemOpen
  \bibfield  {author} {\bibinfo {author} {\bibfnamefont {W.~T.}\ \bibnamefont
  {Jin}}, \bibinfo {author} {\bibfnamefont {N.}~\bibnamefont {Qureshi}},
  \bibinfo {author} {\bibfnamefont {Z.}~\bibnamefont {Bukowski}}, \bibinfo
  {author} {\bibfnamefont {Y.}~\bibnamefont {Xiao}}, \bibinfo {author}
  {\bibfnamefont {S.}~\bibnamefont {Nandi}}, \bibinfo {author} {\bibfnamefont
  {M.}~\bibnamefont {Babij}}, \bibinfo {author} {\bibfnamefont
  {Z.}~\bibnamefont {Fu}}, \bibinfo {author} {\bibfnamefont {Y.}~\bibnamefont
  {Su}},\ and\ \bibinfo {author} {\bibfnamefont {T.}~\bibnamefont
  {Br\"uckel}},\ }\bibinfo {title} {Spiral magnetic ordering of the Eu moments
  in ${\mathrm{EuNi}}_{2}{\mathrm{As}}_{2}$},\ \href
  {https://doi.org/10.1103/PhysRevB.99.014425} {\bibfield  {journal} {\bibinfo
  {journal} {Phys. Rev. B}\ }\textbf {\bibinfo {volume} {99}},\ \bibinfo
  {pages} {014425} (\bibinfo {year} {2019})}\BibitemShut {NoStop}%
\bibitem [{\citenamefont {Bi}\ \emph {et~al.}(2021)\citenamefont {Bi},
  \citenamefont {Nix}, \citenamefont {Dutta}, \citenamefont {Zhao},
  \citenamefont {Alp}, \citenamefont {Zhang}, \citenamefont {Chow},
  \citenamefont {Xiao}, \citenamefont {Liu}, \citenamefont {Cao},\ and\
  \citenamefont {Vohra}}]{bi2021.Eu112Ni}%
  \BibitemOpen
  \bibfield  {author} {\bibinfo {author} {\bibfnamefont {W.}~\bibnamefont
  {Bi}}, \bibinfo {author} {\bibfnamefont {Z.}~\bibnamefont {Nix}}, \bibinfo
  {author} {\bibfnamefont {U.}~\bibnamefont {Dutta}}, \bibinfo {author}
  {\bibfnamefont {J.}~\bibnamefont {Zhao}}, \bibinfo {author} {\bibfnamefont
  {E.~E.}\ \bibnamefont {Alp}}, \bibinfo {author} {\bibfnamefont
  {D.}~\bibnamefont {Zhang}}, \bibinfo {author} {\bibfnamefont
  {P.}~\bibnamefont {Chow}}, \bibinfo {author} {\bibfnamefont {Y.}~\bibnamefont
  {Xiao}}, \bibinfo {author} {\bibfnamefont {Y.-B.}\ \bibnamefont {Liu}},
  \bibinfo {author} {\bibfnamefont {G.-H.}\ \bibnamefont {Cao}},\ and\ \bibinfo
  {author} {\bibfnamefont {Y.~K.}\ \bibnamefont {Vohra}},\ }\bibinfo {title}
  {Microscopic phase diagram of
  $\mathrm{Eu}({\mathrm{Fe}}_{1\ensuremath{-}x}{\mathrm{Ni}}_{x}){\mathrm{As}}_{2}$
  ($x$ = 0,0.04) under pressure},\ \href
  {https://doi.org/10.1103/PhysRevB.103.195135} {\bibfield  {journal} {\bibinfo
   {journal} {Phys. Rev. B}\ }\textbf {\bibinfo {volume} {103}},\ \bibinfo
  {pages} {195135} (\bibinfo {year} {2021})}\BibitemShut {NoStop}%
\bibitem [{\citenamefont {Martinelli}\ \emph {et~al.}(2016)\citenamefont
  {Martinelli}, \citenamefont {Bernardini},\ and\ \citenamefont
  {Massidda}}]{alberto2016}%
  \BibitemOpen
  \bibfield  {author} {\bibinfo {author} {\bibfnamefont {A.}~\bibnamefont
  {Martinelli}}, \bibinfo {author} {\bibfnamefont {F.}~\bibnamefont
  {Bernardini}},\ and\ \bibinfo {author} {\bibfnamefont {S.}~\bibnamefont
  {Massidda}},\ }\bibinfo {title} {The phase diagrams of iron-based
  superconductors: Theory and experiments},\ \href
  {http://www.sciencedirect.com/science/article/pii/S1631070515001267}
  {\bibfield  {journal} {\bibinfo  {journal} {C. R. Phys.}\ }\textbf {\bibinfo
  {volume} {17}},\ \bibinfo {pages} {5 } (\bibinfo {year} {2016})}\BibitemShut
  {NoStop}%
\bibitem [{\citenamefont {Luo}\ and\ \citenamefont {Chen}(2015)}]{luo2015}%
  \BibitemOpen
  \bibfield  {author} {\bibinfo {author} {\bibfnamefont {X.}~\bibnamefont
  {Luo}}\ and\ \bibinfo {author} {\bibfnamefont {X.}~\bibnamefont {Chen}},\
  }\bibinfo {title} {Crystal structure and phase diagrams of iron-based
  superconductors},\ \href
  {https://link.springer.com/content/pdf/10.1007%2Fs40843-015-0022-9.pdf}
  {\bibfield  {journal} {\bibinfo  {journal} {Sci. China Mater.}\ }\textbf
  {\bibinfo {volume} {58}},\ \bibinfo {pages} {77} (\bibinfo {year}
  {2015})}\BibitemShut {NoStop}%
\bibitem [{\citenamefont {Li}\ \emph {et~al.}(2009)\citenamefont {Li},
  \citenamefont {Luo}, \citenamefont {Wang}, \citenamefont {Chen},
  \citenamefont {Ren}, \citenamefont {Tao}, \citenamefont {Li}, \citenamefont
  {Lin}, \citenamefont {He}, \citenamefont {Zhu}, \citenamefont {Cao},\ and\
  \citenamefont {Xu}}]{li2009}%
  \BibitemOpen
  \bibfield  {author} {\bibinfo {author} {\bibfnamefont {L.~J.}\ \bibnamefont
  {Li}}, \bibinfo {author} {\bibfnamefont {Y.~K.}\ \bibnamefont {Luo}},
  \bibinfo {author} {\bibfnamefont {Q.~B.}\ \bibnamefont {Wang}}, \bibinfo
  {author} {\bibfnamefont {H.}~\bibnamefont {Chen}}, \bibinfo {author}
  {\bibfnamefont {Z.}~\bibnamefont {Ren}}, \bibinfo {author} {\bibfnamefont
  {Q.}~\bibnamefont {Tao}}, \bibinfo {author} {\bibfnamefont {Y.~K.}\
  \bibnamefont {Li}}, \bibinfo {author} {\bibfnamefont {X.}~\bibnamefont
  {Lin}}, \bibinfo {author} {\bibfnamefont {M.}~\bibnamefont {He}}, \bibinfo
  {author} {\bibfnamefont {Z.~W.}\ \bibnamefont {Zhu}}, \bibinfo {author}
  {\bibfnamefont {G.~H.}\ \bibnamefont {Cao}},\ and\ \bibinfo {author}
  {\bibfnamefont {Z.~A.}\ \bibnamefont {Xu}},\ }\bibinfo {title}
  {Superconductivity induced by Ni doping in {BaFe$_2$As$_2$} single
  crystals},\ \href {https://doi.org/10.1088/1367-2630/11/2/025008} {\bibfield
  {journal} {\bibinfo  {journal} {New J. Phys.}\ }\textbf {\bibinfo {volume}
  {11}},\ \bibinfo {pages} {025008} (\bibinfo {year} {2009})}\BibitemShut
  {NoStop}%
\bibitem [{\citenamefont {Xing}\ \emph {et~al.}(2015)\citenamefont {Xing},
  \citenamefont {Zhou}, \citenamefont {Xu}, \citenamefont {Li}, \citenamefont
  {Sun}, \citenamefont {Zhang},\ and\ \citenamefont {Shi}}]{xing2015.112Co}%
  \BibitemOpen
  \bibfield  {author} {\bibinfo {author} {\bibfnamefont {X.}~\bibnamefont
  {Xing}}, \bibinfo {author} {\bibfnamefont {W.}~\bibnamefont {Zhou}}, \bibinfo
  {author} {\bibfnamefont {B.}~\bibnamefont {Xu}}, \bibinfo {author}
  {\bibfnamefont {N.}~\bibnamefont {Li}}, \bibinfo {author} {\bibfnamefont
  {Y.}~\bibnamefont {Sun}}, \bibinfo {author} {\bibfnamefont {Y.}~\bibnamefont
  {Zhang}},\ and\ \bibinfo {author} {\bibfnamefont {Z.}~\bibnamefont {Shi}},\
  }\bibinfo {title} {Co-co-doping Effect on Superconducting Properties of
  112-Type {Ca$_{0.8}$La$_{0.2}$FeAs$_2$} Single Crystals},\ \href
  {https://doi.org/10.7566/JPSJ.84.075001} {\bibfield  {journal} {\bibinfo
  {journal} {J. Phys. Soc. Jpn.}\ }\textbf {\bibinfo {volume} {84}},\ \bibinfo
  {pages} {075001} (\bibinfo {year} {2015})}\BibitemShut {NoStop}%
\bibitem [{\citenamefont {Xie}\ \emph {et~al.}(2017)\citenamefont {Xie},
  \citenamefont {Gong}, \citenamefont {Zhang}, \citenamefont {Gu},
  \citenamefont {Huesges}, \citenamefont {Chen}, \citenamefont {Liu},
  \citenamefont {Hao}, \citenamefont {Meng}, \citenamefont {Lu}, \citenamefont
  {Li},\ and\ \citenamefont {Luo}}]{xie2017}%
  \BibitemOpen
  \bibfield  {author} {\bibinfo {author} {\bibfnamefont {T.}~\bibnamefont
  {Xie}}, \bibinfo {author} {\bibfnamefont {D.}~\bibnamefont {Gong}}, \bibinfo
  {author} {\bibfnamefont {W.}~\bibnamefont {Zhang}}, \bibinfo {author}
  {\bibfnamefont {Y.}~\bibnamefont {Gu}}, \bibinfo {author} {\bibfnamefont
  {Z.}~\bibnamefont {Huesges}}, \bibinfo {author} {\bibfnamefont
  {D.}~\bibnamefont {Chen}}, \bibinfo {author} {\bibfnamefont {Y.}~\bibnamefont
  {Liu}}, \bibinfo {author} {\bibfnamefont {L.}~\bibnamefont {Hao}}, \bibinfo
  {author} {\bibfnamefont {S.}~\bibnamefont {Meng}}, \bibinfo {author}
  {\bibfnamefont {Z.}~\bibnamefont {Lu}}, \bibinfo {author} {\bibfnamefont
  {S.}~\bibnamefont {Li}},\ and\ \bibinfo {author} {\bibfnamefont
  {H.}~\bibnamefont {Luo}},\ }\bibinfo {title} {Crystal growth and phase
  diagram of 112-type iron pnictide superconductor
  {Ca$_{1-y}$La$_{y}$Fe$_{1-x}$Ni$_{x}$As$_2$}},\ \href
  {https://doi.org/10.1088/1361-6668/aa7994} {\bibfield  {journal} {\bibinfo
  {journal} {Supercond. Sci. Technol.}\ }\textbf {\bibinfo {volume} {30}},\
  \bibinfo {pages} {095002} (\bibinfo {year} {2017})}\BibitemShut {NoStop}%
\bibitem [{\citenamefont {Raffius}\ \emph {et~al.}(1993)\citenamefont
  {Raffius}, \citenamefont {M\"{o}rsen}, \citenamefont {Mosel}, \citenamefont
  {M\"{u}ller-Warmuth}, \citenamefont {Jeitschko}, \citenamefont
  {Terb\"{u}chte},\ and\ \citenamefont {Vomhof}}]{Raffius1993}%
  \BibitemOpen
  \bibfield  {author} {\bibinfo {author} {\bibfnamefont {H.}~\bibnamefont
  {Raffius}}, \bibinfo {author} {\bibfnamefont {E.}~\bibnamefont {M\"{o}rsen}},
  \bibinfo {author} {\bibfnamefont {B.}~\bibnamefont {Mosel}}, \bibinfo
  {author} {\bibfnamefont {W.}~\bibnamefont {M\"{u}ller-Warmuth}}, \bibinfo
  {author} {\bibfnamefont {W.}~\bibnamefont {Jeitschko}}, \bibinfo {author}
  {\bibfnamefont {L.}~\bibnamefont {Terb\"{u}chte}},\ and\ \bibinfo {author}
  {\bibfnamefont {T.}~\bibnamefont {Vomhof}},\ }\bibinfo {title} {Magnetic
  properties of ternary lanthanoid transition metal arsenides studied by
  M\"{o}ssbauer and susceptibility measurements},\ \href
  {https://doi.org/https://doi.org/10.1016/0022-3697(93)90301-7} {\bibfield
  {journal} {\bibinfo  {journal} {J. Phys. Chem. Solids}\ }\textbf {\bibinfo
  {volume} {54}},\ \bibinfo {pages} {135} (\bibinfo {year} {1993})}\BibitemShut
  {NoStop}%
\end{thebibliography}

%

\end{document}